%% file: 00-main.tex
\documentclass[sigconf]{acmart}
\copyrightyear{2025}
\acmYear{2025}
\setcopyright{rightsretained}

\acmConference[SIGIR '25]{Proceedings of the 48th International ACM SIGIR Conference on Research and Development in Information Retrieval}{July 13--18, 2024}{Padua, Italy.}
\acmBooktitle{Proceedings of the 48th Int'l ACM SIGIR Conference on Research and Development in Information Retrieval (SIGIR '25), July 13--18, 2024, Padua, Italy}

\makeatother

\usepackage{enumitem}
\newlist{inlinelist}{enumerate*}{1}
\setlist*[inlinelist,1]{%
  label=(\roman*),
}
\usepackage{subfigure}
\usepackage{booktabs}
\usepackage{multirow}
\usepackage{array}
\usepackage{makecell}
\usepackage{colortbl}
\usepackage{xcolor}
\usepackage{bbm}
\usepackage{arydshln}


\title{Open-Ended and Knowledge-Intensive Video Question Answering}

\author{Md Zarif Ul Alam}
\affiliation{\institution{University of Massachusetts Amherst}
\country{United States}}
\email{zarifalam@cs.umass.edu}

\author{Hamed Zamani}
\affiliation{\institution{University of Massachusetts Amherst}
\country{United States}}
\email{zamani@cs.umass.edu}

\begin{document}

\begin{abstract}
Video question answering that requires external knowledge beyond the visual content remains a significant challenge in AI systems. While models can effectively answer questions based on direct visual observations, they often falter when faced with questions requiring broader contextual knowledge. To address this limitation, we investigate knowledge-intensive video question answering (KI-VideoQA) through the lens of multi-modal retrieval-augmented generation, with a particular focus on handling open-ended questions rather than just multiple-choice formats. Our comprehensive analysis examines various retrieval augmentation approaches using cutting-edge retrieval and vision language models, testing both zero-shot and fine-tuned configurations. We investigate several critical dimensions: the interplay between different information sources and modalities, strategies for integrating diverse multi-modal contexts, and the dynamics between query formulation and retrieval result utilization. Our findings reveal that while retrieval augmentation shows promise in improving model performance, its success is heavily dependent on the chosen modality and retrieval methodology. The study also highlights the critical role of query construction and retrieval depth optimization in effective knowledge integration. Through our proposed approach, we achieve a substantial 17.5\% improvement in accuracy on multiple choice questions in the KnowIT VQA dataset, establishing new state-of-the-art performance levels.
\end{abstract}

% \keywords{Video Question Answering; Retrieval Augmented Generation; Vision Language Models; Multi-Modal Question Answering}

% <concept_id>10002951.10003317</concept_id>
% <concept_desc>Information systems~Information retrieval</concept_desc>
% <concept_id>10010147.10010257</concept_id>
% <concept_desc>Computing methodologies~Machine learning</concept_desc>

% \ccsdesc[500]{Information systems~Information retrieval}
% \ccsdesc[500]{Computing methodologies~Machine learning}

\maketitle

\input{01-intro}
\input{02-related-work}

\input{03-dataset}

\input{04-problem-statement}

% \section{Experimental Setup}
% \hamed{this can be a subsection if it's too brief.}

\input{05-results}

\input{06-conclusion}

\bibliographystyle{ACM-Reference-Format}
\bibliography{XX-references}

\end{document}

%% file: 01-intro.tex
\section{Introduction}
With the rise of effective vision language models (VLMs), multi-modal question answering tasks have attracted considerable attention. Visual question answering (VQA) \cite{antol2015vqa, goyal2017making} is one of these tasks, where a question is asked about an image. In VQA, the questions can often be answered solely by processing the given image (e.g., `\texttt{what is the color of the table in the image?}'). In knowledge-intensive visual question answering (KI-VQA), on the other hand, researchers study more real-world questions where answering the question requires access to some external knowledge sources, for example, KI-VQA concerns with questions like `\texttt{when was this building [referring to the image] constructed?}'. The knowledge sources studied in KI-VQA are often in the form of structured (or semi-structured) data, such as knowledge bases \cite{marino2019ok,mavex,krisp,kat,unifer,wang2015explicit}. More recently, KI-VQA has been extended to unstructured knowledge sources (e.g., a large collection of text passages) by focusing on passage retrieval \cite{Qu2021KIVQA,Salemi2023ICTIR,Salemi2023MMFID}.

\begin{figure*}[ht]
  % \begin{center}
  \centering
  \includegraphics[width=.9\linewidth]{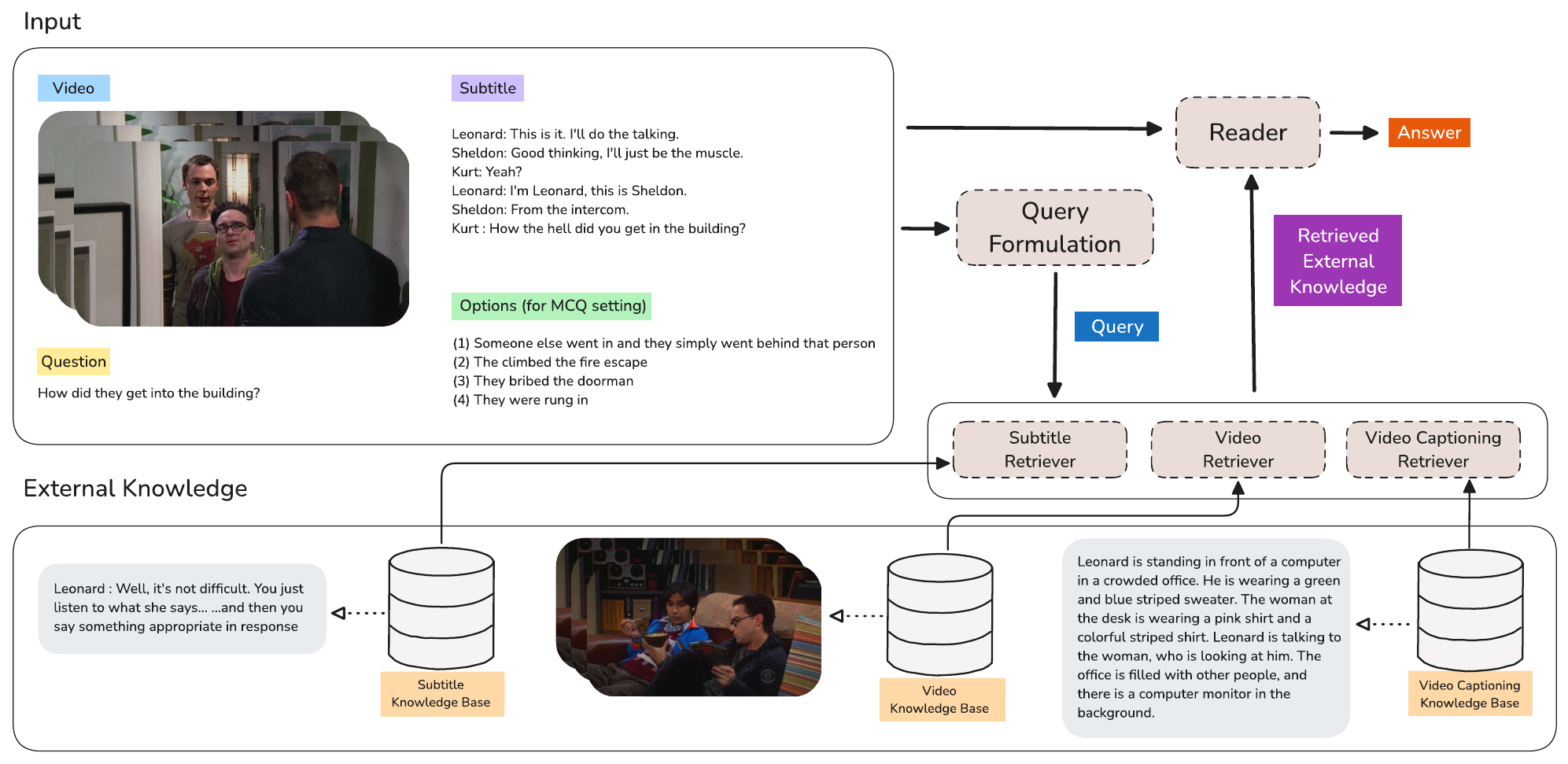}
  \caption{An overview of our multi-modal retrieval augmentation pipeline for KI-VideoQA tasks.}
  \label{fig:method_overview}
  % \end{center}
\end{figure*}

\citet{garcia2020knowit}  expanded KI-VQA to short videos as its visual component.  The task of knowledge-intensive video question answering (KI-VideoQA) concerns with answering questions in the context of a video, where extra information is necessary to accurately answer the question. An example of KI-VideoQA instances is provided in Figure~\ref{fig:method_overview} (top left corner). This task, which is the focus of this work, stands at the intersection of computer vision, information retrieval, and natural language processing. In more detail, KI-VideoQA extends KI-VQA to dynamic scenarios, such as the movement state of objects (e.g., `\texttt{slow}' or `\texttt{fast}'?, `\texttt{put down}' or `\texttt{take away}'?), action repetitions and their transitions over time \citep{jang2017tgif,jang2019video}. To the best of our knowledge, existing work on KI-VideoQA suffers from the following two shortcomings: (1) Prior work assumes that large (vision) language models contain the necessary external information required for answering KI-VideoQA tasks. (2) Prior work on KI-VideoQA  solely focuses on multiple choice questions, which limits the real-world applications of these systems.

We address these two shortcomings as follows: we introduce the first multi-modal retrieval-augmented generation pipeline for KI-VideoQA. This enables us to retrieve unstructured information (textual or visual) from one or more data collection(s) to provide the VLM with the necessary external knowledge. In addition, we extend the KI-VideoQA task to open-ended questions, where no options are provided and the model is expected to generate a free-form textual answer. For this purpose, we adopt and modify established KI-VideoQA benchmarks with multiple choice questions. % that are constructed based on two popular sitcom TV series--The Bing Bang Theory\footnote{\url{https://www.imdb.com/title/tt0898266/}} and Friends.\footnote{\url{https://www.imdb.com/title/tt0108778/}}

Based on these two main contributions, we study various implementations of the developed pipeline using current state-of-the-art retrieval and large VLMs in order to empirically study seven important research questions from an information retrieval perspective.
\begin{itemize}[leftmargin=*]
    \item \textbf{RQ1}: How well do state-of-the-art vision language models perform in KI-VQA tasks? 
\end{itemize}
To answer this question, we evaluate several state-of-the-art VLMs, including commercial models like GPT-4V \cite{achiam2023gpt} and open-source models like Qwen2VL \cite{wang2024qwen2vl} and Intern2VL \cite{wang2024intern2vl} on the KnowIT dataset \cite{garcia2020knowit}. We study these models under different circumstances, including both multiple choice and open-ended questions. We also explore the impact of fine-tuning on open-source VLMs. This research question not only establishes baseline results for our research, but also enables us to better understand the strengths and limitations of VLMs in handling questions that require external knowledge beyond the video content. 

Next, we focus on retrieval-augmented VLMs for KI-VQA. We first explore the impact of different knowledge sources that can be used for retrieval augmentation. In more detail, we address these questions:
\begin{itemize}[leftmargin=*]
    \item \textbf{RQ2}: How does augmenting VLMs with retrieved textual knowledge impact the end-to-end KI-VQA performance? 
    \item \textbf{RQ3}: How does augmenting VLMs with retrieved multi-media knowledge (videos) impact the end-to-end KI-VQA performance?
    \item \textbf{RQ4}: How does retrieval from heterogeneous information sources with multiple modalities (text and video) impact the end-to-end KI-VQA performance? 
\end{itemize}

To answer these research questions, we construct two text corpora for knowledge retrieval. The first corpus consists of subtitles of all videos in the dataset, allowing us to evaluate the impact of past dialogues and narratives in the dataset. We construct the second corpus by automatically generating video captions for each video in the dataset. Video captions often include deep semantic description of video content, serving as a proxy textual representation of the video.  We also explore the collection of all videos in the dataset as a multi-media corpus.  In our experiments, VLMs consume videos frame-by-frame and frames are uniformly sampled at the rate of 1 frame per second. Since these knowledge sources provide complementary information, we also conduct experiments on various combinations of these heterogeneous and multi-modal corpora. For each corpus, we explore three diverse and effective retrieval models, including sparse and dense retrieval models.
We further study:
\begin{itemize}[leftmargin=*]
    \item \textbf{RQ5}: How does our empirical findings transfer to another dataset?
\end{itemize}
By extending our analysis to the KnowIT-X dataset \cite{wu2021transferring}, we assess the generalizability and transferability of the behavior of the developed multi-modal retrieval augmented pipeline for KI-VideoQA. Our analysis also includes transfer learning experiments. 
We finally expand our analysis by exploring:
\begin{itemize}[leftmargin=*]
    \item \textbf{RQ6}: How sensitive are the retrieval-augmented VLMs to the number of retrieved information?
    \item \textbf{RQ7}: How do different query formulations for knowledge retrieval impact the end-to-end performance in KI-VideoQA?
\end{itemize}
In more detail, we explore four different query formulations with and without enriching the input question with the information extracted from the given video. 

In summary, for this underexplored and complex task that demands expertise across various disciplines, we find it crucial to conduct thorough empirical analyses. These analyses should address multiple facets of the problem, including the selection of knowledge sources, the retrieval model's capability to effectively retrieve information from these sources, the VLM's proficiency in utilizing the retrieved (multi-modal) context, and the formulation of effective queries for the task. Such a comprehensive approach lays the groundwork for developing robust multi-modal RAG systems for KI-VideoQA tasks. Besides these important lessons, our contributions advance state-of-the-art accuracy by over 17.5\% (from 65.2 to 76.7\%) for multiple choice questions on the KnowIT dataset.\footnote{Note that prior work only studied multiple choice questions.} We will soon release our implementation publicly for improving the reproducibility of the results.

%% file: 02-related-work.tex
\section{Related Work}
This section reviews prior work on video understanding and question answering as well as knowledge-intensive visual QA.
% \hamed{note to self: decide later whether a section on RAG should be added or not.}

\paragraph{\textbf{Video Question Answering}}
VideoQA has evolved significantly from simple visual recognition to complex temporal and knowledge-based reasoning tasks. This evolution can be traced through three distinct stages \cite{xiao2024videoqa}. Initially, approaches combined CNNs for visual features with RNNs for text processing \cite{jang2017tgif, xu2017video}, focusing primarily on visual content understanding. The field then progressed to more sophisticated architectures like Vision Transformers \cite{VisionTransformer} and BERT-based models that leveraged cross-modal pre-training \cite{lei2018tvqa, fu2021violet} in the second stage.

The current era features VLMs that combine visual encoders like CLIP \cite{radford2021learning} or EVA-CLIP \cite{sun2023eva} with powerful language models such as LLaMA \cite{touvron2023llama} and Vicuna \cite{chiang2023vicuna}. Notable examples include VideoChat \cite{li2023videochat}, Video-LLaVA \cite{lin2023video}, LLaMA-VID \cite{li2025llama}, InternVL \cite{chen2024internvl}, and QwenVL \cite{bai2023qwenvl}. These models have made significant contributions, which demonstrate strong capabilities in video understanding and temporal reasoning. Commercial models like GPT-4V \cite{achiam2023gpt} and Gemini \cite{team2023gemini} have further pushed the boundaries, achieving near-human performance on some visual understanding tasks.

While these models excel at standard VideoQA tasks, recent studies \cite{xiao2024can} reveal concerning limitations, e.g., models can often answer questions correctly with irrelevant or even no video inputs, suggesting reliance on language priors rather than true visual understanding. Additionally, current models struggle with temporal reasoning and content ordering, highlighting the gap between model capabilities and human-like video comprehension \cite{bagad2023test}.

This work extends the current state-of-the-art VideoQA models to knowledge-intensive questions through retrieval augmentation, an area of research that is yet underexplored.

\paragraph{\textbf{Knowledge-Intensive Visual Question Answering}}
KI-VQA emerged to address the fundamental limitation of standard VQA systems--the constraint of inferring answers solely from training data. Given the finite nature of any training dataset, the knowledge scope of traditional VQA systems remains inherently limited. KI-VQA approaches aim to overcome this by incorporating external knowledge sources into the question answering process. Unlike this work, KI-VQA models mostly focus on images.

Early systems focused on common sense-enabled VQA, where models were augmented with basic world knowledge. These initial approaches either created specialized image-focused knowledge bases with template questions \cite{zhu2015building} or extracted information from general-purpose knowledge bases like DBpedia \cite{wu2016ask, auer2007dbpedia} to enhance VQA accuracy.
Several datasets have been introduced to study different aspects of knowledge integration.  KB-VQA \cite{wang2015explicit} and FVQA \cite{wang2017fvqa} focus on questions generated from templates or knowledge base facts. R-VQA \cite{lu2018r} studies relational fact reasoning, requiring models to understand relationships between entities. OK-VQA \cite{marino2019ok} contains free-form factoid questions without explicit knowledge annotations, requiring models to use an external knowledge base for answering natural language questions about images.

Most early KI-VQA datasets imposed significant constraints on question formulation, either through template-based generation or direct extraction from knowledge bases. This limited their ability to capture the complexity and diversity of real-world questions. KnowIT VQA \cite{garcia2020knowit} and KnowIT-X VQA \cite{wu2021transferring} datasets extended knowledge-based reasoning to video understanding. They were the first datasets to explore external knowledge questions in video content, with questions freely proposed by qualified workers to study both knowledge integration and temporal coherence. They are used in this research and are introduced in more detail in Section~\ref{sec:data}. While significant progress has been made in image-based KI-VQA \cite{Qu2021KIVQA,Salemi2023ICTIR,Salemi2023MMFID,mavex,krisp,kat,unifer}, extending these approaches to video understanding (i.e., KI-VideoQA) presents unique challenges that are understudied. For instance, it is still unclear what and how external information should be used for effective development of KI-VideoQA systems. This paper bridges these gaps.

%% file: 03-dataset.tex
\section{Datasets}
\label{sec:data}
To empirically study our research questions, we adopt two recent knowledge-intensive video question answering datasets that contain multi-choice questions. To the best of our knowledge, there is no publicly available dataset for generating free-form answers in response to knowledge-intensive questions about a video. Therefore, we construct another version of each dataset for free-form question answering tasks. In more detail, we simply use the textual description of the correct option as the reference (ground-truth) answer and do not provide the options to the QA models that are supposed to generate free-form answers. The datasets we adopt are KnowIT VQA \cite{garcia2020knowit} and KnowIT-X VQA \cite{wu2021transferring}. Both of them are constructed from popular TV sitcoms and feature questions that require integrating visual, textual, temporal, and external knowledge understanding.

\paragraph{\textbf{KnowIT VQA}}
KnowIT VQA \cite{garcia2020knowit} contains video clips from the TV show ``The Big Bang Theory'' and comprises 24,282 human-annotated question-answer pairs. The dataset was constructed from 207 episodes, resulting in 12,087 video clips of approximately 20 seconds each. Each clip is accompanied by subtitles and human-written questions that span different types: visual-based (22\%), textual-based (12\%), temporal-based (4\%), and knowledge-based (62\%). Importantly, each question-video pair includes an annotated knowledge snippet that captures the external information required to answer the question correctly. The models do not use this annotated knowledge snippet to simulate real-world question answering tasks over videos.

\paragraph{\textbf{KnowIT-X VQA}}
KnowIT-X VQA \cite{wu2021transferring} extends the methodology used by KnowIT VQA to another popular sitcom TV show, ``Friends''. It contains 21,412 question-video pairs from 202 episodes, divided into 12,176 video clips. Similar to its predecessor, data samples include video clips, subtitles, questions, four options (one correct), and an annotated snippet as the knowledge information.
% Similarly, the models in our experiments do not have access to this manually annotated knowledge snippet and are expected to find relevant knowledge from various sources that will be described in each experiment.

% \hamed{note to self: later check whether needed to mention information about retrieval corpora here.}

% \begin{table}[t]
% \caption{Statistics of the KnowIT and KnowIT-X datasets.}
% \label{tab:dataset_stats}
% % \small
% \begin{tabular}{lrr}
% \toprule
%  & \textbf{KnowIT} & \textbf{KnowIT-X} \\
% \midrule
% % \multicolumn{3}{l}{Dataset Scale} \\
% \# Episodes & 207 & 202 \\
% \# Scenes & 2,472 & 2,565 \\
% \# Short video clips & 12,087 & 12,176 \\
% \# Video-Question-Answer Tuples & 24,282 & 21,412 \\
% \midrule
% % \multicolumn{3}{l}{Average length (\# tokens)} \\
% Average question length & 7.49 & 7.65 \\
% Average subtitles length & 56.79 & 39.06 \\
% Average ground-truth answer length & 4.54 & 2.30 \\
% % Wrong Ans. & 4.13 & 2.06 \\ 
% \bottomrule
% \end{tabular}
% \end{table}

%% file: 04-problem-statement.tex
\section{Multi-Modal RAG for KI-VideoQA}
\label{sec:method}
In this work, we address the task of knowledge-intensive open-ended video question answering. Formally, given an input video $V$, a question $q$ about the video, and a corpus or knowledge source $\mathcal{K}$, the task is to answer the question. Note that in this task, questions cannot be merely answered by the given video; answering the question requires external knowledge. An example of this task is presented in Figure~\ref{fig:method_overview} (top left corner).

\paragraph{\textbf{Inputs}}
Let $V = [f_1, f_2, ..., f_t]$ be a video sequence of $t$ frames, where each frame $f_t \in \mathbb{R}^{h \times w \times 3}$ represents an RGB image with the height of $h$ and width of $w$ pixels. Each video is accompanied by subtitles ($S$) between the start and end timestamp for that video. Each question $q$ is a natural language sequence of $l$ tokens ${w_1, w_2, \cdots, w_l}$.

\paragraph{\textbf{Knowledge Sources}}
A knowledge source $\mathcal{K}$ is a collection of $m$ information items: $\mathcal{K} = \{k_1, k_2, ..., k_m\}$, where each information item may contain information needed to answer some questions. Each information item can be either textual (e.g., unstructured text documents) or visual (e.g., videos). This paper studies these and even multi-modal knowledge source situations which is a mixture of both textual and visual information items. 

\paragraph{\textbf{Output}}
Our work encompasses two distinct question answering tasks: (1) \textbf{Multiple Choice Questions}: Given a set of candidate answers $\mathcal{A} = {a^1, a^2, ..., a^K}$ where $K$ is typically equal to 4, the task is to select one correct answer $a^* \in \mathcal{A}$. (2) \textbf{Open-Ended Questions}: The expected output is a free-form text with the length of $p$ as the answer: $a = {y_1, y_2, \cdots, y_p}$, where each $y_i \in \mathcal{V}$ is a token from the vocabulary $\mathcal{V}$.

% \begin{enumerate}
%     \item \textbf{Multiple Choice Questions}: Given a set of candidate answers $\mathcal{A} = {a^1, a^2, ..., a^K}$ where $K$ is typically equal to 4, the task is to select one correct answer $a^* \in \mathcal{A}$. 
% %     The model outputs a probability distribution over the candidates:
% % \begin{equation}
% % P(a^k|\mathcal{V}, q, \mathcal{K}) \quad \text{for } k \in {1,...,K}
% % \end{equation}

%     \item \textbf{Open-Ended Questions}: The expected output is a free-form text with the length of $p$ as the answer: $a = {y_1, y_2, \cdots, y_p}$, where each $y_i \in \mathcal{V}$ is a token from the vocabulary $\mathcal{V}$.
% \end{enumerate}

Even though multiple choice questions for KI-VideoQA has been briefly explored in \citet{garcia2020knowit}, prior work has not studied retrieval-augmented solutions. Moreover, to the best of our knowledge, no prior work explored open-ended question answering in KI-VideoQA.

% \subsection{Multi-Modal RAG for KI-VideoQA}
One approach to answer questions is to feed the given question and video (or its subtitle) to a VLM for answer generation. This approach has been studied in \cite{garcia2020knowit,wu2021transferring}. However, we believe that this approach cannot perform well for many knowledge-intensive questions. If the LLM has observed and learned the ``knowledge'' required for answering the question, this approach is likely to generate a correct answer, however, this is not always possible. For instance, this approach would not perform well in non-stationary data situations where new videos and knowledge sources have been produced after the VLM training. Therefore, this work presents the first attempt to apply ideas from the retrieval-augmented text generation (RAG) literature \cite{reml,rag} to KI-VideoQA.

In the following, we introduce a generic framework and discuss various implementations of it in each experiment. The framework, as depicted in Figure~\ref{fig:method_overview}, consists of three main components: query formulation, multi-modal knowledge retrieval, and a VLM for information synthesis and answer generation. The framework processes input in the form of a text question, video frames, and additional information about the video if available such as associated subtitles, utilizing external knowledge bases to generate accurate answers in both multiple-choice and open-ended settings. In the following, we describe various implementations of each of these components studied in this paper.

\subsection{Query Formulation}
The effectiveness of retrieval heavily depends on query formulation. We explore several query formulation strategies:
\begin{enumerate}[leftmargin=*]
\item \textbf{Question-only}: Using the question text $q$ as the query:
% \begin{equation}
% q_{raw} = q
% \end{equation}

\item \textbf{Question + Options}: For multiple choice questions, we can concatenate the question with all provided options as the query. This query formulation cannot be applied to open-ended answer generation, as options are not available to the model in this setting.
% concatenating the question with all possible options:
% \begin{equation}
%     q_{opt} = q \oplus \{a^1, a^2, ..., a^K\}
% \end{equation}
% where $\oplus$ denotes concatenation and $\{a^1, ..., a^K\}$ are the candidate answers.

\item \textbf{Question + Subtitle}: Enriching the query with the video subtitle, by concatenating the question with the video subtitle.
% \hamed{In your equation you mentioned subtitle at time $t$. What is $t$ here?} 
% relevant subtitle context:
% \begin{equation}
%     q_{sub} = q \oplus s_t
% \end{equation}
% where $s_t$ represents the subtitle at timestamp $t$.

\item \textbf{LLM-Based Query Rewriting}: Rewriting the question using a VLM that takes the video, subtitle and question, and is prompted to rewrite the question for higher quality retrieval. \footnote{Detailed prompts available in source code}
% Applying transformations to the base query using the inputs with a VLM:
% \begin{equation}
%     q_{trans} = T(q)
% \end{equation}
% where $T(\cdot)$ is the transformation that reformulates the query for better retrieval.

\end{enumerate}

\subsection{Multi-Modal Knowledge Retrieval}

Given a formulated query $q^*$ obtained from the last component, we perform retrieval over a diverse set of multi-modal knowledge sources as follows:
\begin{enumerate}[leftmargin=*]
\item \textbf{Subtitle Retrieval}: This retrieval model takes a collection of videos as a knowledge source and uses their subtitles to construct a text document for every video in the collection. Therefore, this knowledge source basically contains historical dialogues in the video collection and enables the system to leverage conversational context and spoken information that may not be visually apparent.

% \begin{equation}
% \mathcal{R}s = R_s(q*, \mathcal{K}_s)
% \end{equation}

% \begin{equation}
%     \mathcal{R}_c = R_c(q_*, \mathcal{K}_c)
% \end{equation}

\item \textbf{Video Caption Retrieval}: Developing effective video retrieval models is challenging and one approach is to turn the video into a textual description through video captioning. Therefore, each video will be represented by a single text document, automatically generated using a large vision-language model. We use Qwen2-VL-2B \cite{wang2024qwen2vl} in zero-shot setting for this purpose. These captions serve as an intermediate representation bridging visual and textual modalities. To provide an insight into what a video caption may contain, we provide one example in Figure~\ref{fig:method_overview} (bottom right corner).

% This corpus provides dense semantic descriptions of video content.

\item \textbf{Video Retrieval}: Knowledge retrieval can be done directly to the collection of videos. We do video retrieval using the generated captions and finding the corresponding video of the retrieved captions. This component helps in understanding visual context, actions, and temporal relationships that may be crucial for answering some question.

% \begin{equation}
%     \mathcal{R}_c = R_c(q_*, \mathcal{K}_c)
% \end{equation}

\end{enumerate}

Each retriever component operates independently and can be implemented using various retrieval architectures, such as sparse or dense retrieval models. Each returns a ranked list of information items with the highest relevance score, given the formulated query $q^*$. We implement and evaluate three distinct retrieval models:
\begin{itemize}[leftmargin=*]
    \item \textbf{BM25 \cite{robertson1994some}}: A sparse retrieval model rooted in classical probabilistic models. We use Elasticsearch's implementation of BM25. We use the default BM25 parameters (i.e., $b=0.75$ and $k_1=1.2$). This approach can only be employed for subtitle and video caption retrieval.

     \item \textbf{NV-Embed-v2 \cite{lee2024nv}}: State-of-the-art dense retrieval model which was ranked No. 1 on the Massive Text Embedding Benchmark \cite{muennighoff-etal-2023-mteb} as of January 5, 2025 with an impressive score of 72.31 across 56 text embedding tasks. The model is built on the Mistral-7B-v0.1 architecture and produces embeddings with a dimension of 4096. 
     % They used several new ideas: a two-staged instruction tuning method to enhance the accuracy of both retrieval and non-retrieval tasks, used Latent-attention mechanism to allows the LLM to attend to latent vectors, resulting in improved pooled embedding output. 
     % \hamed{it would be nice if you include more information about this. how many parameters? what is especial about it? for instance, a model that is trained using knowledge distillation on the x and y dataset.}
    
    \item \textbf{Stella \cite{zhang2024jasper}}: We use another state-of-the-art dense retrieval model, stella\_en\_400M\_v5.\footnote{Available at \url{https://huggingface.co/dunzhang/stella_en_400M_v5}.} This model only has 400M parameters and encodes queries and documents into 1024-dimensional dense vectors. Compared to other similar performing embedding models, both the number of parameters and encoded vector dimension are very small; for example, NV-Embed-v2 \cite{lee2024nv}, bge-en-icl \cite{li2024making}, and e5-mistral-7b-instruct \cite{wang2022text, wang2023improving, wang2024multilingual} have 7B parameters, and their vector dimensions are 4096. The deployment and application of these larger models in industry were hampered by their vector dimensions and numerous parameters, making it too slow for practical use. Stella uses an innovative distillation technique to achieve high performance while maintaining a smaller footprint. In our experiments we saw 11x speed improvement for a fixed batch size during encoding, compared to NV-Embed-v2.

% This makes it a compact yet powerful dense retrieval model that addresses the industry deployment challenges of large vector dimensions.
    
    % The model encodes queries and documents into 1024-dimensional dense vectors. Note that Stella has shown improved performance compared to commonly used dense retrieval models, such as DPR, ANCE, and TAS-B \cite{}. \hamed{I added the last sentence. Please find a good citation for it, so people don't question your choice.}

\end{itemize}

% For all retrieval models, we maintain a consistent evaluation setup. 

% \hamed{maybe better to highlight retrieval models and also the fact that we consider heterogenuous information at some point.}

\subsection{Augmented Answer Generation}

The final component is a VLM that generates the answer. It takes the original inputs (video, question, subtitle and options for the MCQ setting) and retrieved knowledge from each of the retrievers. The model utilizes heterogenuous information both from the input and retrieved knowledge. We utilize Qwen2-VL with two billion parameters \cite{wang2024qwen2vl} as our primary reader model, which serves as the foundation for answering both multiple choice and open-ended questions. Qwen2-VL processes the input video frames, question, and retrieved knowledge simultaneously to generate answers. For video frame processing, we sample frames uniformly at 1 FPS and encode them using Qwen2VL's native visual encoder. Each frame height and width is resized into 224 pixels before feeding to the VLM.
% \hamed{more information. what LLMs? any specific prompt? zero-shot or fine-tuning.}

\subsection{Fine-Tuning}
This section describes the method used for fine-tuning the Qwen2VL 2B model. The model is initialized using the pre-trained weights available on HuggingFace.\footnote{Available at \url{https://huggingface.co/Qwen/Qwen2-VL-2B-Instruct}.} We use the Adam optimizer with weight decay (AdamW) \cite{loshchilov2017decoupled} for fine-tuning with a learning rate of 1e-5. We use the cross-entropy loss function with a batch size of 1 due to the high memory requirement of processing videos by VLMs. To minimize the impact of gradient fluctuation, we update model parameters with 50 gradient accumulation step, resulting in an effective batch size of 50. The model uses Flash Attention 2 \cite{dao2023flashattention2fasterattentionbetter} for better acceleration and memory efficiency, especially when processing multiple videos. We use the training portion of each dataset for fine-tuning the model.

% \hamed{here describe your model training. loss funciton. optimization. etc.}

% Model: Qwen2VL
% Model Size: 2B parameters
% Pre-trained Model: The model is initialized with the pre-trained weights from Qwen/Qwen2-VL-2B-Instruct
% Optimizer: AdamW (Adam with Weight Decay)
% Learning Rate: 1e-5
% Loss Function: Cross-Entropy Loss
% Batch Size: 1 (due to the high memory requirements of processing images and videos)
% Gradient Accumulation Steps: 50 (Gradient accumulation is used to simulate a larger batch size by accumulating gradients over multiple forward passes before performing a backward pass and updating the model parameters. This helps in managing memory constraints while still allowing for effective training.)
% Flash Attention: The model uses flash\_attention\_2 for better acceleration and memory efficiency, especially when processing multiple images or videos.
% Training data: train partition of the datasets

%% file: 05-results.tex
\section{Results and Discussion}
In this section, we present our empirical results to study seven research questions. All experiments were conducted using 1 NVIDIA A100 GPU with 80GB memory. We use half-precision (fp16) computations to optimize memory usage and processing speed.

\paragraph{\textbf{Evaluation Metrics}} We evaluate our models using a wide range of metrics. Following prior work \cite{garcia2020knowit,wu2021transferring}, we use accuracy to measure model's effectiveness for multiple choice questions. For open-ended questions, we use ROUGE-1, ROUGE-L \cite{lin-2004-rouge}, BLEU-1, BLEU-4 \cite{papineni2002bleu}, and term overlap F1 (TO F1) \cite{Kong2013} for measuring various aspects of lexical overlap between the generated answer and the reference ground truth answer. We also use METEOR \cite{banerjee2005meteor} and BERTScore \cite{zhang2019bertscore} for measuring semantic similarity. This diverse set of metrics allows us to evaluate different aspects of answer quality, from lexical accuracy to semantic appropriateness. All metrics are computed on the test sets of each dataset.

To measure statistically significant improvements, we use a two-tailed paired t-test. Due to the binary nature of accuracy metric per query, we use McNemar test for this metric.
We report significant improvements where p-value is less than 0.05.

\begin{table*}[t]
\centering
\caption{The results obtained by state-of-the-art VLMs for both multiple choice and open-ended questions on the KnowIT VQA dataset, under both zero-shot and fine-tuned settings. Current best reported performance on the KnowITVQA dataset is also shown in the table. Boldface indicates highest number in the column.}
\resizebox{1.8\columnwidth}{!}{%
\begin{tabular}{llccccccccc}
\toprule
\multirow{2}{*}{\textbf{Model}} & \multirow{2}{*}{\textbf{Setting}} & \multicolumn{1}{c}{\textbf{MCQ}} & & \multicolumn{7}{c}{\textbf{Open-Ended Questions}} \\
\cmidrule{3-3}\cmidrule{5-11}
 & & \textbf{\% Accuracy} & & \textbf{ROUGE-1} & \textbf{ROUGE-L} & \textbf{METEOR} & \textbf{BLEU-1} & \textbf{BLEU-4} & \textbf{TO F1} & \textbf{BERTScore} \\
\midrule
ROCK \cite{garcia2020knowit} & Fine-tuned & 65.20 && - & - & -& -& -& -& -   \\
\midrule 
\multirow{2}{*}{Qwen2VL 2B} & Zero-shot & 52.01 && 0.1266&	0.1156&	0.1471&	0.0882&	0.0163&	0.1086&	0.0777 \\
 & Fine-tuned & \textbf{67.34} && \textbf{0.3481}&	\textbf{0.3408}&	\textbf{0.2162}&	\textbf{0.3022}&	\textbf{0.0761}&	\textbf{0.2770}&	\textbf{0.4477} \\
\midrule 
\multirow{2}{*}{InternVL2-2B} & Zero-shot & 50.78 && 0.1450&	0.1345&	0.1548&	0.1027&	0.0200&	0.1265&	0.1177 \\
 & Fine-tuned & 67.17&& 0.3397&	0.3331&	0.2098&	0.2942&	0.0724&	0.2702&	0.4328 \\
\midrule 
GPT-4V & Zero-shot & 65.77&& 0.1327&	0.1225&	0.1685&	0.0813&	0.0172&	0.1313&	0.0493 \\ 
% \cline{4-10}
 % & Fine-tuned & x & \multicolumn{7}{c|}{x} \\
\bottomrule
\end{tabular}
}
\label{tab:vlm_results}
\end{table*}

\begin{table*}[t]
\centering
\caption{The results obtained by a retrieval augmented Qwen2VL 2B model \cite{wang2024qwen2vl} on the KnowIT VQA dataset using three diverse retrieval models (i.e., BM25, Stella, and NV-Embed-v2). We use three different knowledge sources for retrieval augmentation (i.e., video subtitles, automatically generated video captions, and videos themselves). Boldface indicates highest number in the column. Best performing zero-shot results are highlighted using underline. The superscript $^\triangle$ indicates improvement compared to the same VLM under the same experimental setting without retrieval augmentation. $^\blacktriangle$ means improvements are statistically significant.}
\resizebox{1.8\columnwidth}{!}{%
\begin{tabular}{llllllllllll}
\toprule
{\textbf{Retrieval}} & \textbf{Retrieval} & \multirow{2}{*}{\textbf{Setting}} & \multicolumn{1}{c}{\textbf{MCQ}} && \multicolumn{7}{c}{\textbf{Open-Ended Questions}} \\
\cmidrule{4-4}\cmidrule{6-12}
\textbf{Corpus} & \textbf{Model} &&  \textbf{\% Accuracy} & & \textbf{ROUGE-1} & \textbf{ROUGE-L} & \textbf{METEOR} & \textbf{BLEU-1} & \textbf{BLEU-4} & \textbf{TO F1} & \textbf{BERTScore} \\
\midrule
\multirow{6}{*}{\rotatebox{270}{Subtitles}} & \multirow{2}{*}{BM25}  & Zero-shot & 44.89 && 0.1282$^\triangle$ & 0.1169$^\triangle$ & 0.1391 & 0.0931$^\blacktriangle$ & 0.0180$^\blacktriangle$ & 0.1082 & 0.1061$^\blacktriangle$ \\
% \cline{4-12}
 & & Fine-tuned & 70.27$^\blacktriangle$  && 0.3586$^\triangle$ & 0.3501$^\triangle$ & 0.2260$^\blacktriangle$ & 0.3117$^\triangle$ & 0.0805$^\blacktriangle$ & 0.3020$^\blacktriangle$ & 0.4522$^\triangle$ \\
\cmidrule{2-12}
 & \multirow{2}{*}{Stella}  & Zero-shot & 47.81  && 0.1375$^\triangle$ & 0.1259$^\triangle$ & 0.1514$^\blacktriangle$ & 0.0992$^\blacktriangle$ & 0.0199$^\blacktriangle$ & 0.1183$^\blacktriangle$ & 0.1134$^\blacktriangle$ \\
% \cline{4-12}
 & & Fine-tuned & 72.34$^\blacktriangle$  && 0.3842$^\blacktriangle$ & 0.3763$^\blacktriangle$ & 0.2437$^\blacktriangle$ & 0.3320$^\blacktriangle$ & 0.0889$^\blacktriangle$ & 0.3245$^\blacktriangle$ & 0.4561$^\blacktriangle$ \\
\cmidrule{2-12}
 & \multirow{2}{*}{NV-Embed-v2}  & Zero-shot & 51.67  && \underline{0.1434}$^\blacktriangle$ & \underline{0.1316}$^\blacktriangle$ & \underline{0.1583}$^\blacktriangle$ & \underline{0.1009}$^\blacktriangle$ & \underline{0.0201}$^\blacktriangle$ & \underline{0.1280}$^\blacktriangle$ & \underline{0.1156}$^\blacktriangle$ \\
% \cline{4-12}
 & & Fine-tuned & \textbf{74.84$^\blacktriangle$}  && \textbf{0.3950$^\blacktriangle$} & \textbf{0.3862$^\blacktriangle$} & \textbf{0.2511$^\blacktriangle$} & \textbf{0.3429$^\blacktriangle$} & \textbf{0.0904$^\blacktriangle$} & \textbf{0.3394$^\blacktriangle$} & \textbf{0.4656$^\blacktriangle$} \\
\midrule

 \multirow{6}{*}{\rotatebox{270}{Video Captions}} & \multirow{2}{*}{BM25} & Zero-shot & 32.23 && 0.0868 & 0.0789 & 0.1009 & 0.0627 & 0.0112 & 0.0614 & 0.0192 \\
% \cline{4-12}
& & Fine-tuned & {66.71} & &  0.3428 & 0.3363 & 0.2149 & 0.2992 & 0.0753 & 0.2765 & 0.4421 \\
\cmidrule{2-12}
& \multirow{2}{*}{Stella} & Zero-shot & 40.83  & & 0.1012 & 0.0906 & 0.1163 & 0.0724 & 0.0128 & 0.0766 & 0.0563 \\
% \cline{4-12}
& & Fine-tuned & 65.90 & &  {0.3473} & 0.3415$^\triangle$ & {0.2200$^\triangle$} & {0.3050$^\triangle$} & {0.0767$^\triangle$} & {0.2795$^\triangle$} & {0.4449} \\
\cmidrule{2-12}
& \multirow{2}{*}{NV-Embed-v2} & Zero-shot & 40.03 & &  0.0930 & 0.0841 & 0.1098 & 0.0672 & 0.0119 & 0.0717 & 0.0449 \\
% \cline{4-12}
 & & Fine-tuned & 66.41  && 0.3421 & 0.3354 & 0.2150 & 0.2991 & 0.0762$^\triangle$ & 0.2749 & 0.4487$^\triangle$ \\
\midrule\midrule

\multirow{6}{*}{\rotatebox{270}{Videos}} & \multirow{2}{*}{BM25} & Zero-shot & \underline{53.16}$^\blacktriangle$ && 0.1264&	0.1151&	0.1466&	0.0878&	0.0161&	0.1085&	0.0761\\

% \cline{4-12}
 & & Fine-tuned & 65.81 && 0.3488$^\triangle$&	0.3407&	0.2165$^\triangle$&	0.3017&	0.0755&	0.2791$^\triangle$&	0.4477\\
\cmidrule{2-12}
& \multirow{2}{*}{Stella} & Zero-shot & 50.69&& 		0.1154&	0.1046&	0.1317&	0.0817&	0.0148&	0.0911&	0.1034$^\triangle$ \\
% \cline{4-12}
& & Fine-tuned & 63.78&	& 	0.3362&	0.3285 &	0.2049&	0.2875&	0.0716&	0.2648&	0.4350 \\
\cmidrule{2-12}
& \multirow{2}{*}{NV-Embed-v2} & Zero-shot & 50.65&	& 	0.1259&	0.1147&	0.1431&	0.0880&	0.0159&	0.0991&	0.0789$^\triangle$ \\
% \cline{4-12}
& & Fine-tuned & 58.95&& 		0.3302&	0.3220&	0.2016&	0.2820&	0.0700&	0.2558&	0.4322 \\
\bottomrule
\end{tabular}
}
\label{tab:diff_corpora}
\end{table*}

\begin{table*}[t]
\caption{The results obtained by a retrieval augmented Qwen2VL 2B model \cite{wang2024qwen2vl} using multiple heterogeneous corpora as knowledge sources on the KnowIT VQA dataset. NV-Embed-v2 (our best performing retrieval model) is used for information retrieval. Boldface indicates highest number in the column. Best performing zero-shot results are highlighted using underline. The superscript $^\blacktriangle$ indicates improvement compared to the same VLM under the same experimental setting without retrieval augmentation.}
\centering
\resizebox{1.8\columnwidth}{!}{%
\begin{tabular}{lllcccccccc}
\toprule
 \textbf{Retrieval} & \multirow{2}{*}{\textbf{Setting}} & \multicolumn{1}{c}{\textbf{MCQ}} & \multicolumn{7}{c}{\textbf{Open-Ended Questions}} \\
\cmidrule{3-3} \cmidrule{5-11}
\textbf{Corpora} & & \textbf{\% Accuracy} && \textbf{ROUGE-1} & \textbf{ROUGE-L} & \textbf{METEOR} & \textbf{BLEU-1} & \textbf{BLEU-4} & \textbf{TO F1} & \textbf{BERTScore} \\
\midrule
Subtitles \&   & Zero-shot & 46.67 && \underline{0.1225} & \underline{0.1117} & 0.1468 & \underline{0.0854} & \underline{0.0164}$^\triangle$ & \underline{0.1069} & \underline{0.0762} \\
% \cline{4-12}
Video Captions & Fine-tuned & \textbf{75.60}$^\blacktriangle$ && \textbf{0.3947}$^\blacktriangle$ & \textbf{0.3874}$^\blacktriangle$ & \textbf{0.2494}$^\blacktriangle$ & \textbf{0.3445}$^\blacktriangle$ & \textbf{0.0903}$^\blacktriangle$ & \textbf{0.3388}$^\blacktriangle$ & 0.4678$^\blacktriangle$ \\
\midrule
 Subtitles \&  & Zero-shot & \underline{56.92}$^\blacktriangle$&&		0.1219&	0.1106&	\underline{0.1513}$^\triangle$ &	0.0826&	0.0157&	0.1049&	0.0698 \\
Videos & Fine-tuned & 73.31$^\blacktriangle$&&		0.3855$^\blacktriangle$&	0.3775$^\blacktriangle$&	0.2419$^\blacktriangle$&	0.3363$^\blacktriangle$&	0.0873$^\blacktriangle$&	0.3267$^\blacktriangle$&	0.4653$^\blacktriangle$ \\
\midrule
 Videos \&  & Zero-shot & 40.61&	&	0.0981&	0.0882&	0.1155&	0.0681&	0.0117&	0.074&	0.0274 \\
% \cline{4-12}
Video Captions & Fine-tuned & 65.26&&		0.3363&	0.3294&	0.2080&	0.2931&	0.0727&	0.2668&	0.4417 \\
\midrule
 Subtitles \& Videos \& & Zero-shot & 48.50 & & 0.1089 & 0.0984 & 0.1316 & 0.0743 & 0.0136 & 0.0869 & 0.0483 \\
% \cline{4-12}
 Video Captions  & Fine-tuned & 73.78$^\blacktriangle$ && 0.3867$^\blacktriangle$ & 0.3788$^\blacktriangle$ & 0.2470$^\blacktriangle$ & 0.3375$^\blacktriangle$ & 0.0886$^\blacktriangle$ & 0.3289$^\blacktriangle$ & \textbf{0.4692}$^\blacktriangle$ \\
\bottomrule
\end{tabular}
}
\label{tab:heterogenuous}
\end{table*}

\begin{table*}[t]
\centering
\caption{The results obtained by a (retrieval augmented) Qwen2VL 2B model \cite{wang2024qwen2vl} for different retrieval corpora on the \textit{KnowIT-X} VQA dataset. NV-Embed-v2 (our best performing retrieval model) is used for information retrieval. Boldface indicates highest number in the column. Best performing zero-shot results are highlighted using underline.}
\resizebox{1.8\columnwidth}{!}{%
\begin{tabular}{llccccccccc}
\toprule
\textbf{Retrieval} & \multirow{2}{*}{\textbf{Setting}} & \multicolumn{1}{c}{\textbf{MCQ}} & & \multicolumn{7}{c}{\textbf{Open-Ended Questions}} \\
\cmidrule{3-3}\cmidrule{5-11}
\textbf{Corpus} & & \textbf{\% Accuracy} & & \textbf{ROUGE-1} & \textbf{ROUGE-L} & \textbf{METEOR} & \textbf{BLEU-1} & \textbf{BLEU-4} & \textbf{TO F1} & \textbf{BERTScore} \\
\midrule 
\multirow{3}{*}{None} & Zero-shot & 54.88 && 0.0995&	0.0952&	0.1431&	0.0768&	0.0177&	0.1115&	0.0534 \\
& Fine-tuned & 64.42&& 0.2894&	0.2870&	0.1719&	0.2483&	0.0664&	0.2608&	0.4989\\
& Transfer learning & 55.13&&		0.2894&	0.287&	0.1719&	0.2483&	0.0664&	0.2608&	0.4989\\\midrule

\multirow{3}{*}{Subtitles} & Zero-shot & 60.35 && 0.1268& 0.121&	0.1763&	0.0939&	0.0217&	0.1414&	0.0986 \\
& Fine-tuned & 74.29 && 0.5472&0.5423&	0.3646&	0.5051&	0.1559&	0.5083&	0.6419 \\
& Transfer learning & 70.71 && 0.4032&	0.399&	0.234&	0.3432&	0.0852&	0.3826&	0.5365 \\\midrule

\multirow{3}{*}{Video Captions} & Zero-shot & 48.40 && 0.0574&	0.0546&	0.0922&	0.048&	0.0104&	0.0583&	0.0016\\
& Fine-tuned & 62.29 && 0.4652&	0.4604&	0.3251&	0.4359&	0.1432&	0.4227&	0.6178 \\
& Transfer learning & 52.81 && 0.2386&	0.2368&	0.1447&	0.2031&	0.0544&	0.213&	0.4746\\\midrule

\multirow{3}{*}{Videos} & Zero-shot & 51.74 && 0.096&	0.0911&	0.1384&	0.076&	0.0176&	0.1017&	0.0561 \\
& Fine-tuned & 62.82 && 0.4591&	0.4541&	0.3093&	0.4265&	0.1357&	0.4157&	0.6067 \\
& Transfer learning & 57.6 && 0.2833&	0.28&	0.1624&	0.2421&	0.0628&	0.2572&	0.4786 \\\midrule

\multirow{3}{*}{Subtitles \& Video Captions} & Zero-shot & 49.80&& 0.0912&	0.0871&	0.1426&	0.0673&	0.0144&	0.1056&	0.0416 \\
& Fine-tuned & \textbf{74.44}&& \textbf{0.5479}&	\textbf{0.5426}&	\textbf{0.3693}&	\textbf{0.5084}&	\textbf{0.1570}&	\textbf{0.5147}&	\textbf{0.6479} \\
& Transfer learning & 68.58&& 0.4009&	0.3971&	0.2349&	0.3399&	0.0838&	0.3754&	0.5258 \\\midrule

\multirow{3}{*}{Subtitles \& Videos} & Zero-shot & 59.15&& 0.1016&	0.0968&	0.1564&	0.0726&	0.0155&	0.1196&	0.0522  \\
& Fine-tuned & 73.38&& 0.5387&	0.5333&	0.3625&	0.5007&	0.1554&	0.5030&	0.6364 \\
& Transfer learning & 70.04&& 0.3854&	0.3828&	0.2217&	0.3292&	0.0811&	0.3622&	0.5217 \\\midrule

\multirow{3}{*}{Videos \& Video Captions} & Zero-shot & 42.49&& 0.0513&	0.0487&	0.0857&	0.0431&	0.009&	0.0472&	0.0213 \\
& Fine-tuned & 60.50&& 0.4505&	0.4460&	0.3100&	0.4212&	0.1345&	0.4099&	0.5998 \\
& Transfer learning & 57.26&&	0.2366&	0.2347&	0.1403&	0.1987&	0.0523&	0.2126&	0.4629 \\\midrule

\multirow{3}{*}{Subtitles \& Videos \& Video Captions} & Zero-shot & 52.08&& 0.0756&	0.0718&	0.1231&	0.0558&	0.0113&	0.086&	0.0106 \\
& Fine-tuned & 72.19&& 0.5382&	0.5333&	0.3647&	0.5006&	0.1553&	0.5028&	0.6407 \\
& Transfer learning & 68.93&& 0.3866&	0.3818&	0.2262&	0.3341&	0.0818&	0.363&	0.5244 \\
\bottomrule
\end{tabular}
}
\label{tab:knowitx}
\end{table*}

\begin{table*}[t]
\centering
\caption{The results obtained by a retrieval augmented Qwen2VL 2B model \cite{wang2024qwen2vl} using video subtitles as the knowledge source on the KnowIT VQA dataset. NV-Embed-v2 (our best performing retrieval model) is used for information retrieval. Boldface indicates highest number in the column. Best performing zero-shot results are highlighted using underline.}
\resizebox{1.6\columnwidth}{!}{%
\begin{tabular}{llccccccccc}
\toprule
 \multirow{2}{*}{\textbf{Setting}} & \multirow{2}{*}{\textbf{k}} & \multicolumn{1}{c}{\textbf{MCQ}} & & \multicolumn{7}{c}{\textbf{Open-Ended Questions}} \\
\cmidrule{3-3}\cmidrule{5-11}
 & & \textbf{\% Accuracy} & & \textbf{ROUGE-1} & \textbf{ROUGE-L} & \textbf{METEOR} & \textbf{BLEU-1} & \textbf{BLEU-4} & \textbf{TO F1} & \textbf{BERTScore} \\
\hline
\multirow{5}{*}{{Zero-shot}} & 0 & 52.01 && 0.1266&	0.1156&	0.1471&	0.0882&	0.0163&	0.1086&	0.0777 \\
 & 2 & 54.21 & & 0.1527 & 0.1413 & 0.1658 & 0.1081 & 0.0226 & 0.1370 & 0.1248 \\
&{5}  & 54.60 && \underline{0.1536} & \underline{0.1426} & \underline{0.1689} & \underline{0.1082} & \underline{0.0228} & \underline{0.1375} & \underline{0.1290} \\
 &{10}  & 51.67 && 0.1434 & 0.1316 & 0.1583 & 0.1009 & 0.0201 & 0.1280 & 0.1156 \\
&{20} & \underline{55.23} && 0.1225 & 0.1127 & 0.1383 & 0.0876 & 0.0171 & 0.1042 & 0.0946 \\
\midrule

\multirow{5}{*}{{Fine-tuned}} & 0 & {67.34} && {0.3481}&	{0.3408}&	{0.2162}&	{0.3022}&	{0.0761}&	{0.2770}&	{0.4477} \\
 &{2}  & 70.73 && 0.3764 & 0.3686 & 0.239 & 0.3273 & 0.0861 & 0.3151 & 0.4632 \\
% \cline{4-13}
% \cline{5-13}
& {5}  & 74.54 & & 0.3978 & 0.3894 & 0.2525 & 0.3466 & 0.0920 & 0.3411 & \textbf{0.4785} \\
% \cline{4-13}
% \cline{5-13}
&{10}  & 74.84 && 0.3950 & 0.3862 & 0.2511 & 0.3429 & 0.0904 & 0.3394 & 0.4656 \\
% \cline{4-13}
% \cline{5-13}
& {20}  & \textbf{74.96} & & \textbf{0.4036} & \textbf{0.3962} & \textbf{0.2556} & \textbf{0.3502} & \textbf{0.0915} & \textbf{0.3518} & 0.4677 \\
\midrule
\end{tabular}
}
\label{tab:topk-comparison}
\end{table*}

\begin{table*}[t]
\centering
\caption{The results obtained by a retrieval augmented Qwen2VL 2B model \cite{wang2024qwen2vl} for four different query formulation strategies on the KnowIT VQA dataset. Video subtitles are used as the knowledge source. Boldface indicates highest number in the column. Best performing zero-shot results are highlighted using underline.}
\resizebox{1.8\columnwidth}{!}{%
\small
\begin{tabular}{llllllllllll}
\toprule
\multirow{2}{*}{\textbf{Query}} & \textbf{Retrieval} & \multirow{2}{*}{\textbf{Setting}} & \multicolumn{1}{c}{\textbf{MCQ}} && \multicolumn{7}{c}{\textbf{Open-Ended Questions}} \\
\cmidrule{4-4}\cmidrule{6-12}
 & \textbf{Model} &&  \textbf{\% Accuracy} & & \textbf{ROUGE-1} & \textbf{ROUGE-L} & \textbf{METEOR} & \textbf{BLEU-1} & \textbf{BLEU-4} & \textbf{TO F1} & \textbf{BERTScore} \\
\midrule

\multirow{6}{*}{\rotatebox{270}{question-only}} & \multirow{2}{*}{BM25} & Zero-shot & 32.91 && 0.1181 & 0.1076 & 0.1258 & 0.0875 & 0.0161 & 0.0908 & 0.1019 \\
 & & Fine-tuned & 67.01 && 0.3580 & 0.3498 & 0.2224 & 0.3105 & 0.0790 & 0.2878 & 0.4494 \\
\cmidrule{2-12}
 & \multirow{2}{*}{Stella} & Zero-shot & 44.30 && 0.1261 & 0.1155 & 0.1377 & 0.0924 & 0.0180 & 0.1003 & 0.1096 \\
 & & Fine-tuned & 70.61 && 0.3755 & 0.3671 & 0.2346 & 0.3243 & 0.0841 & 0.3156 & 0.4536 \\
\cmidrule{2-12}
 & \multirow{2}{*}{NV-Embed-v2} & Zero-shot & 44.64 && 0.1332 & 0.1217 & 0.1442 & 0.0964 & 0.0191 & 0.1044 & 0.1118 \\
 & & Fine-tuned & 70.81 && 0.3830 & 0.3761 & 0.2419 & 0.3335 & 0.0877 & 0.3235 & \textbf{0.4667} \\
\midrule
\multirow{6}{*}{\rotatebox{270}{question \& options}}& \multirow{2}{*}{BM25} & Zero-shot & 39.05 && - &- &- &- &- &- &-  \\
 & & Fine-tuned & 76.23 && - &- &- &- &- &- &- \\
\cmidrule{2-12}
 & \multirow{2}{*}{Stella} & Zero-shot & 46.71 && - &- &- &- &- &- &- \\
& & Fine-tuned & 73.90 && - &- &- &- &- &- &- \\
\cmidrule{2-12}
 & \multirow{2}{*}{NV-Embed-v2} & Zero-shot & 47.64 && - &- &- &- &- &- &- \\
 & & Fine-tuned & \textbf{76.75} && - &- &- &- &- &- &- \\
\midrule
\multirow{6}{*}{\rotatebox{270}{question \& subtitle}} & \multirow{2}{*}{BM25} & Zero-shot & 44.89 && 0.1282 & 0.1169 & 0.1391 & 0.0931 & 0.0180 & 0.1082 & 0.1061 \\
 & & Fine-tuned & 70.27 && 0.3586 & 0.3501 & 0.2260 & 0.3117 & 0.0805 & 0.3020 & 0.4522 \\
\cmidrule{2-12}
 & \multirow{2}{*}{Stella} & Zero-shot & 47.81& & 0.1375 & 0.1259 & 0.1514 & 0.0992 & 0.0199 & 0.1183 & 0.1134 \\
 & & Fine-tuned & 72.34 && 0.3842 & 0.3763 & 0.2437 & 0.3320 & 0.0889 & 0.3245 & 0.4561 \\
\cmidrule{2-12}
 & \multirow{2}{*}{NV-Embed-v2} & Zero-shot & \underline{51.67} && 0.1434 & 0.1316 & 0.1583 & 0.1009 & 0.0201 & 0.1280 & 0.1156 \\
 & & Fine-tuned & 74.84 & &\textbf{0.3950} & \textbf{0.3862} & \textbf{0.2511} & \textbf{0.3429} & \textbf{0.0904} & \textbf{0.3394} & 0.4656 \\
\midrule 
\multirow{6}{*}{\rotatebox{270}{question rewriting}} & \multirow{2}{*}{BM25} & Zero-shot & 36.63&& 0.1363&	0.1260 &	0.1484&	0.0999&	0.0217&	0.1153&	0.1108 \\
 & & Fine-tuned &  62.30 && 0.3506&	0.3446&	0.2220 &	0.3046&	0.0786&	0.2931&	0.4473\\
\cmidrule{2-12}
 & \multirow{2}{*}{Stella} & Zero-shot & 45.10 && \underline{0.1485} &	\underline{0.1373} &	\underline{0.1660} &	\underline{0.1058} &	\underline{0.0220} &	\underline{0.1287} &	\underline{0.1232} \\
 & & Fine-tuned & 65.01 && 0.3441&	0.3374&	0.2213&	0.2947&	0.0757&	0.2827&	0.4383 \\
\cmidrule{2-12}
 & \multirow{2}{*}{NV-Embed-v2} & Zero-shot & 47.39 && 0.1259&	0.1147&	0.1431&	0.0880 &	0.0159&	0.0991&	0.0789 \\
 & & Fine-tuned & 64.21 && 0.3409 &	0.3312 &	0.2190 &	0.2921 &	0.0760 &	0.2863 &	0.4323 \\

\bottomrule
\end{tabular}
}
\label{tab:diff_queries}
\end{table*}

\paragraph{\textbf{RQ1: How well do state-of-the-art vision language models perform in KI-VQA tasks?}}
To answer this question, we evaluate several state-of-the-art VLMs, including commercial models like GPT-4V \cite{achiam2023gpt} and open-source models like Qwen2VL \cite{wang2024qwen2vl} and InternVL2 \cite{wang2024intern2vl}. We study these models under different circumstances, including both multiple choice and open-ended questions and both zero-shot and fine-tuned settings for open-source VLMs.  The results are presented in Table~\ref{tab:vlm_results}. We also add the current best performance \cite{garcia2020knowit} in the table. Unsurprisingly, the results suggest that fine-tuning both open-source VLMs substantially improve their performance on the KnowIT VQA dataset. The results also suggest that the GPT-4V is the best performing zero-shot VLM for multiple choice question answering. However, the zero-shot InternVL2 model significantly performs better than GPT-4V in the open-ended setting, in terms of ROUGE-1, ROUGE-L, BLEU-1, BLEU-4, and BERTScore. That being said, the fine-tuned Qwen2VL model demonstrates the highest performance across all metrics. This model achieves an accuracy of over 67\% for multiple choice questions. Questions in this setting have four options, meaning that a random selector would achieve a 25\% accuracy. These results not only establish baseline performance for our research, but also enables us to better understand the strengths and limitations of VLMs in handling questions that require external knowledge beyond the video content.

Due to the strong fine-tuned performance of Qwen2VL in Table~\ref{tab:vlm_results}, we perform all retrieval augmentation experiments using this VLM.

\paragraph{\textbf{RQ2: How does augmenting VLMs with retrieved textual knowledge impact the end-to-end KI-VQA performance?}}
To address this research question, we construct two text corpora for knowledge retrieval as described in Section~\ref{sec:method}. The first corpus consists of subtitles of all videos in the dataset. The second corpus was constructed through automatic captioning of the videos. Video captions often include deep semantic description of video content, serving as a proxy textual representation of the video. For each corpus, we employ three diverse and effective retrieval models (both term matching and dense retrieval models). See the retrieval model details in Section~\ref{sec:method}. In all experiments in this research question, 10 documents are retrieved for augmentation. We report the results in Table~\ref{tab:diff_corpora} (i.e., the top two sections related to subtitles and video captions). 
% , allowing us to evaluate the impact of past dialogues and narratives in the dataset.
Our investigation of subtitle-based retrieval augmentation reveals several important findings. The neural retriever NV-Embed-v2 consistently outperforms other approaches across all metrics, achieving 74.84\% MCQ accuracy and stronger open-ended response scores compared to BM25 and Stella.  Notably, the relatively low absolute scores on open-ended metrics (term overlap F1 scores ranging 0.3020-0.3394) highlight that generating accurate free-form responses remains challenging even with retrieval augmentation. This suggests that improving response generation quality, rather than just retrieval effectiveness, may be a crucial direction for future work.

% The superscript $\triangle$ indicates performance improvement compared to the results obtained by the same VLM without retrieval augmentation (i.e., Qwen2VL 2B results in Table~\ref{tab:vlm_results}).
% Therefore, as is evident from the results, retrieving from the subtitles corpus leads to statistically significant \hamed{?} performance gain in nearly all settings.

Analyzing the results from the video caption-based retrieval augmentation also highlights some important findings. First, augmenting with video captions underperform subtitles in every single experimental setting. The main reason for this observation is that the external knowledge required to answer most questions in the KnowIT dataset is not captured by the video captions. Another reason could be due to the failure of retrieval models in retrieving relevant video captions, compared to subtitles. The second main observation is that retrieval augmentation with video captions often performs even worse than the baseline results presented in Table~\ref{tab:vlm_results} (see Qwen2VL 2B results). For multiple choice questions, video caption-based augmentation deteriorates the results in every case. Underperforming the baseline model in this case means that the VLM cannot accurately and robustly disregard irrelevant information presented to them in retrieval augmentation. For open-ended questions, there are few cases with minor improvements which are negligible. Therefore, the nature of the textual knowledge source and the retrievability of information can substantially impact the behavior of multi-modal retrieval-augmented models in KI-VQA.

\paragraph{\textbf{RQ3: How does augmenting VLMs with retrieved multi-media knowledge (videos) impact the end-to-end KI-VQA performance?}}
VLMs have demonstrated different capabilities in integrating and consuming multi-media content. This research question help us better understand in what modality information needs to be presented to the current VLM technology for effective answer generation. In our experiments, VLMs consume videos frame-by-frame and frames are uniformly sampled at the rate of 1 frame per second. The model details are presented in Section~\ref{sec:method}. The last section of Table~\ref{tab:diff_corpora} reports the video-based retrieval augmentation results. It is worth noting that we observe that multi-modal retrieval models perform poorly for video retrieval, thus we use video subtitles to represents the videos for text-based retrieval; however, their corresponding videos are fed to the VLM for answer generation. 

Interestingly, video-based retrieval augmentation shows notably weaker performance compared to subtitle-based augmentation for open-domain questions.  It is worth highlighting that we could only use one retrieved video because of GPU memory availability. In the fine-tuned setting, even the best performing retriever (BM25) achieves only 65.81\% MCQ accuracy, which is significantly lower than subtitle-based retrieval (74.84\%) and slightly below caption-based retrieval (66.71\%). This performance gap is consistent across all open-ended metrics. However, video-based model demonstrate the strongest zero-shot results for multiple choice questions.

Given this observation, we believe that developing memory efficient VLMs that can consume more videos as context can potentially lead to the best performing multi-modal retrieval augmented systems. Moreover, since this model performs better under the zero-shot setting (compared to the baseline), future work can focus on improving fine-tuning optimizations for video context.

\paragraph{\textbf{RQ4: How does retrieval from heterogeneous information sources with multiple modalities (text and video) impact the end-to-end KI-VQA performance?}}
After examining individual modalities for retrieval augmentation, we investigate whether combining different knowledge sources can further enhance KI-VideoQA performance. We experiment with various combinations of our three knowledge corpora: subtitles (text), video captions (text), and video content (multi-media). For all experiments, we use NV-Embed-v2 as our retriever given its strong performance in single-modality settings, and employ subtitle-enriched queries.

Our results in Table~\ref{tab:heterogenuous} reveal several interesting patterns in multi-modal retrieval. The combination of subtitles and video captions leads to the best fine-tuning performance among all combinations in terms of nearly all metrics (except for BERTScore), reaching 75.6\% MCQ accuracy and the highest scores across all open-ended metrics. Comparing to the results presented in Table~\ref{tab:diff_corpora}, we observe that textual information from different sources (subtitles and captions) complement each other effectively, with captions potentially providing additional semantic context to the already available information in subtitles which is mostly about dialogues and narratives in the video.

Surprisingly, incorporating video retrieval consistently leads to performance degradation in fine-tuned settings. The subtitle-video combination performs worse than subtitle \& caption, and the video \& caption combination  shows the poorest performance (65.26\% MCQ accuracy). Even using all three modalities performs slightly worse than subtitle \& caption alone, achieving 73.78\% MCQ accuracy. 

These findings from fine-tuned experiments indicate that while combining different types of textual information (subtitles and captions) is beneficial, current video retrieval strategy might have introduced noise rather than helpful context. This suggests a need for better video retrieval techniques that can more effectively complement text-based knowledge sources. That said, the highest zero-shot result was achieved by the subtitles \& captions setting, suggesting that zero-shot VLMs can take advantage of multi-media context, but this advantage is lost after fine-tuning.

\paragraph{\textbf{RQ5: How does our empirical findings transfer to another dataset?}}

To assess the generalizability and transferability of our findings, we evaluate our approaches on the KnowIT-X dataset \cite{wu2021transferring}. We explore three experimental settings: zero-shot, fine-tuning, and transfer learning (where we use the model fine-tuned on KnowIT and evaluate it on KnowIT-X). Our results in Table~\ref{tab:knowitx} show several important patterns that largely align with our findings on the KnowIT dataset:

First, subtitle-based retrieval augmentation remains the most effective approach when only one of the knowledge sources are used for retrieval augmentation. When fine-tuned, subtitle retrieval achieves the highest performance (74.29\% MCQ accuracy and consistently stronger open-ended metrics) compared to video captions (62.29\%) and videos (62.82\%). This confirms our earlier observation about the importance of dialogue and narrative information captured in subtitles. Second, combining different knowledge sources shows similar patterns - the subtitle \& video caption combination achieves the strongest performance (74.44\% MCQ accuracy), while incorporating video retrieval tends to slightly degrade performance. This is consistent with our findings on the KnowIT dataset.

In terms of transfer learning, we observe that models pre-trained on the KnowIT dataset can transfer reasonably well to the KnowIT-X dataset. All metrics across MCQ and Open-Ended setting shows that it is significantly better than zero-shot performance but it has some performance degradation compared to the fine-tuned ones. For instance, with subtitle retrieval, transfer learning achieves 70.71\% MCQ accuracy compared to 60.35\% in zero-shot and 74.29\% with fine-tuning. The relatively strong transfer performance suggests that the knowledge integration strategies learned on one TV show can generalize to another, despite differences in content and style. 

% \hamed{write this when the results are ready.} 

\paragraph{\textbf{RQ6: How sensitive are the retrieval-augmented VLMs to the number of retrieved information?}}
To measure model sensitivity to the number of retrieved documents fed to the VLM, we performed an experiments by increasing the number of documents from 2 to 20 for both zero-shot and fune-tuned settings. In this experiments, for the sake of space, we solely focus on our best performing retrieval model NV-Embed-v2 and our best performing open-source VLM, i.e., Qwen2VL 2B. We use the subtitle-enriched questions as the search query. The results are presented in Table~\ref{tab:topk-comparison}. According to the table, the fine-tuned performance keeps increasing for all metrics except for BERTScore as we feed more retrieved documents to the VLM context. Thus the highest performance is achieved when 20 documents is used for augmentation. The model with 5 documents achieves the highest BERTScore. The zero-shot model has a very different behavior. The open-ended performance generally declines if we use more than five documents. This suggests that the zero-shot VLM is not well trained for taking advantage of more documents, since this behavior changes after fine-tuning. That said, the highest MCQ performance is achieved when 20 documents is used. However, the MCQ accuracy does not consistently increase with more documents, i.e., there is a drop in 10 documents. This demonstrates the unpredictable behavior of the zero-shot VLM in consuming long context.

\paragraph{\textbf{RQ7: How do different query formulations for knowledge retrieval impact the end-to-end performance in KI-VideoQA?}}
We study four different query formulations for KI-VideoQA as detailed in Section~\ref{sec:method}. This analysis was conducted on the video subtitles retrieval corpus, as it shows the strongest performance in our previous research questions. The results can be found in Table~\ref{tab:diff_queries}. For the MCQ task, the options-enriched query formulation proves to be the most effective strategy, achieving the highest accuracy across all retrievers (up to 76.747\% with NV-Embed-v2). This suggests that the additional context from answer options helps retrieve more relevant information. However, this advantage is limited to MCQ settings only.

% \hamed{this section will be edited once the results are complete.}
In open-ended scenarios, where $q_{opt}$ is not applicable, subtitle-enriched queries ($q_{sub}$) consistently outperform both raw questions ($q_{raw}$) and transformed queries ($q_{trans}$). With NV-Embed-v2, $q_{sub}$ achieves better scores across all metrics (e.g., ROUGE-1: 0.395 vs 0.383 for $q_{raw}$). Surprisingly, query transformation ($q_{trans}$) performs worst among all strategies in both MCQ and open-ended settings, suggesting that current transformation methods may be introducing noise rather than helpful context.
These findings indicate that while option-based queries are optimal for MCQ tasks, subtitle-enriched queries provide the best general-purpose strategy, especially for open-ended VideoQA where answer options are unavailable. The consistent performance patterns across different retrievers suggest these findings are robust to retrieval architecture choice. We also therefore use subtitle-enriched queries $q_{sub}$ for the rest of our experiments.

%% file: 06-conclusion.tex
\section{Conclusions and Future Directions}

% \hamed{future work: long videos / other domains beyond tv / embodid stuff / attribution}

This paper presents the first comprehensive study of multi-modal retrieval augmentation for knowledge-intensive video question answering. Our extensive experiments across two datasets demonstrate that retrieval augmentation significantly improves KI-VideoQA performance, with subtitle-based retrieval being particularly effective. The combination of subtitles with automatically generated video captions yields the best results. However, our work reveals important limitations in current video retrieval strategies. Directly retrieving video content tends to degrade performance compared to text-based approaches, highlighting the need for more sophisticated video retrieval techniques.

We intend to develop more efficient and effective video retrieval methods specifically designed for KI-VideoQA, as well as extending retrieval augmentation to longer-form videos beyond TV show clips. There are also opportunities to explore embodied question answering scenarios where physical interaction with the environment is required. Additionally, improving attribution and explanation capabilities could help users better understand how retrieved knowledge influences answers.

%% file: 00-main.bbl

\begin{thebibliography}{57}

%%% the Web address.

\ifx \showCODEN    \undefined \def \showCODEN     #1{\unskip}     \fi
\ifx \showDOI      \undefined \def \showDOI       #1{#1}\fi
\ifx \showISBNx    \undefined \def \showISBNx     #1{\unskip}     \fi
\ifx \showISBNxiii \undefined \def \showISBNxiii  #1{\unskip}     \fi
\ifx \showISSN     \undefined \def \showISSN      #1{\unskip}     \fi
\ifx \showLCCN     \undefined \def \showLCCN      #1{\unskip}     \fi
\ifx \shownote     \undefined \def \shownote      #1{#1}          \fi
\ifx \showarticletitle \undefined \def \showarticletitle #1{#1}   \fi
\ifx \showURL      \undefined \def \showURL       {\relax}        \fi
% invisible to TeX
\providecommand\bibfield[2]{#2}
\providecommand\bibinfo[2]{#2}
\providecommand\natexlab[1]{#1}
\providecommand\showeprint[2][]{arXiv:#2}

\bibitem[Antol et~al\mbox{.}(2015)]%
        {antol2015vqa}
\bibfield{author}{\bibinfo{person}{Stanislaw Antol}, \bibinfo{person}{Aishwarya Agrawal}, \bibinfo{person}{Jiasen Lu}, \bibinfo{person}{Margaret Mitchell}, \bibinfo{person}{Dhruv Batra}, \bibinfo{person}{C~Lawrence Zitnick}, {and} \bibinfo{person}{Devi Parikh}.} \bibinfo{year}{2015}\natexlab{}.
\newblock \showarticletitle{Vqa: Visual question answering}. In \bibinfo{booktitle}{\emph{Proceedings of the IEEE international conference on computer vision}}. \bibinfo{pages}{2425--2433}.
\newblock


\bibitem[Auer et~al\mbox{.}(2007)]%
        {auer2007dbpedia}
\bibfield{author}{\bibinfo{person}{S{\"o}ren Auer}, \bibinfo{person}{Christian Bizer}, \bibinfo{person}{Georgi Kobilarov}, \bibinfo{person}{Jens Lehmann}, \bibinfo{person}{Richard Cyganiak}, {and} \bibinfo{person}{Zachary Ives}.} \bibinfo{year}{2007}\natexlab{}.
\newblock \showarticletitle{Dbpedia: A nucleus for a web of open data}. In \bibinfo{booktitle}{\emph{international semantic web conference}}. Springer, \bibinfo{pages}{722--735}.
\newblock


\bibitem[Bagad et~al\mbox{.}(2023)]%
        {bagad2023test}
\bibfield{author}{\bibinfo{person}{Piyush Bagad}, \bibinfo{person}{Makarand Tapaswi}, {and} \bibinfo{person}{Cees~GM Snoek}.} \bibinfo{year}{2023}\natexlab{}.
\newblock \showarticletitle{Test of time: Instilling video-language models with a sense of time}. In \bibinfo{booktitle}{\emph{Proceedings of the IEEE/CVF Conference on Computer Vision and Pattern Recognition}}. \bibinfo{pages}{2503--2516}.
\newblock


\bibitem[Bai et~al\mbox{.}(2023)]%
        {bai2023qwenvl}
\bibfield{author}{\bibinfo{person}{Jinze Bai}, \bibinfo{person}{Shuai Bai}, \bibinfo{person}{Shusheng Yang}, \bibinfo{person}{Shijie Wang}, \bibinfo{person}{Sinan Tan}, \bibinfo{person}{Peng Wang}, \bibinfo{person}{Junyang Lin}, \bibinfo{person}{Chang Zhou}, {and} \bibinfo{person}{Jingren Zhou}.} \bibinfo{year}{2023}\natexlab{}.
\newblock \bibinfo{title}{Qwen-VL: A Versatile Vision-Language Model for Understanding, Localization, Text Reading, and Beyond}.
\newblock
\newblock
\showeprint[arxiv]{2308.12966}~[cs.CV]


\bibitem[Banerjee and Lavie(2005)]%
        {banerjee2005meteor}
\bibfield{author}{\bibinfo{person}{Satanjeev Banerjee} {and} \bibinfo{person}{Alon Lavie}.} \bibinfo{year}{2005}\natexlab{}.
\newblock \showarticletitle{METEOR: An automatic metric for MT evaluation with improved correlation with human judgments}. In \bibinfo{booktitle}{\emph{Proceedings of the acl workshop on intrinsic and extrinsic evaluation measures for machine translation and/or summarization}}. \bibinfo{pages}{65--72}.
\newblock


\bibitem[Chen et~al\mbox{.}(2024)]%
        {chen2024internvl}
\bibfield{author}{\bibinfo{person}{Zhe Chen}, \bibinfo{person}{Jiannan Wu}, \bibinfo{person}{Wenhai Wang}, \bibinfo{person}{Weijie Su}, \bibinfo{person}{Guo Chen}, \bibinfo{person}{Sen Xing}, \bibinfo{person}{Muyan Zhong}, \bibinfo{person}{Qinglong Zhang}, \bibinfo{person}{Xizhou Zhu}, \bibinfo{person}{Lewei Lu}, {et~al\mbox{.}}} \bibinfo{year}{2024}\natexlab{}.
\newblock \showarticletitle{Internvl: Scaling up vision foundation models and aligning for generic visual-linguistic tasks}. In \bibinfo{booktitle}{\emph{Proceedings of the IEEE/CVF Conference on Computer Vision and Pattern Recognition}}. \bibinfo{pages}{24185--24198}.
\newblock


\bibitem[Chiang et~al\mbox{.}(2023)]%
        {chiang2023vicuna}
\bibfield{author}{\bibinfo{person}{Wei-Lin Chiang}, \bibinfo{person}{Zhuohan Li}, \bibinfo{person}{Zi Lin}, \bibinfo{person}{Ying Sheng}, \bibinfo{person}{Zhanghao Wu}, \bibinfo{person}{Hao Zhang}, \bibinfo{person}{Lianmin Zheng}, \bibinfo{person}{Siyuan Zhuang}, \bibinfo{person}{Yonghao Zhuang}, \bibinfo{person}{Joseph~E Gonzalez}, {et~al\mbox{.}}} \bibinfo{year}{2023}\natexlab{}.
\newblock \showarticletitle{Vicuna: An open-source chatbot impressing gpt-4 with 90\%* chatgpt quality}.
\newblock \bibinfo{journal}{\emph{See https://vicuna. lmsys. org (accessed 14 April 2023)}} \bibinfo{volume}{2}, \bibinfo{number}{3} (\bibinfo{year}{2023}), \bibinfo{pages}{6}.
\newblock


\bibitem[Dao(2023)]%
        {dao2023flashattention2fasterattentionbetter}
\bibfield{author}{\bibinfo{person}{Tri Dao}.} \bibinfo{year}{2023}\natexlab{}.
\newblock \bibinfo{title}{FlashAttention-2: Faster Attention with Better Parallelism and Work Partitioning}.
\newblock
\newblock
\showeprint[arxiv]{2307.08691}~[cs.LG]
\urldef\tempurl%
\url{https://arxiv.org/abs/2307.08691}
\showURL{%
\tempurl}


\bibitem[Dosovitskiy et~al\mbox{.}(2021)]%
        {VisionTransformer}
\bibfield{author}{\bibinfo{person}{Alexey Dosovitskiy}, \bibinfo{person}{Lucas Beyer}, \bibinfo{person}{Alexander Kolesnikov}, \bibinfo{person}{Dirk Weissenborn}, \bibinfo{person}{Xiaohua Zhai}, \bibinfo{person}{Thomas Unterthiner}, \bibinfo{person}{Mostafa Dehghani}, \bibinfo{person}{Matthias Minderer}, \bibinfo{person}{Georg Heigold}, \bibinfo{person}{Sylvain Gelly}, \bibinfo{person}{Jakob Uszkoreit}, {and} \bibinfo{person}{Neil Houlsby}.} \bibinfo{year}{2021}\natexlab{}.
\newblock \showarticletitle{An Image is Worth 16x16 Words: Transformers for Image Recognition at Scale}. In \bibinfo{booktitle}{\emph{International Conference on Learning Representations}}.
\newblock


\bibitem[Fu et~al\mbox{.}(2021)]%
        {fu2021violet}
\bibfield{author}{\bibinfo{person}{Tsu-Jui Fu}, \bibinfo{person}{Linjie Li}, \bibinfo{person}{Zhe Gan}, \bibinfo{person}{Kevin Lin}, \bibinfo{person}{William~Yang Wang}, \bibinfo{person}{Lijuan Wang}, {and} \bibinfo{person}{Zicheng Liu}.} \bibinfo{year}{2021}\natexlab{}.
\newblock \showarticletitle{Violet: End-to-end video-language transformers with masked visual-token modeling}.
\newblock \bibinfo{journal}{\emph{arXiv preprint arXiv:2111.12681}} (\bibinfo{year}{2021}).
\newblock


\bibitem[Garcia et~al\mbox{.}(2020)]%
        {garcia2020knowit}
\bibfield{author}{\bibinfo{person}{Noa Garcia}, \bibinfo{person}{Mayu Otani}, \bibinfo{person}{Chenhui Chu}, {and} \bibinfo{person}{Yuta Nakashima}.} \bibinfo{year}{2020}\natexlab{}.
\newblock \showarticletitle{KnowIT VQA: Answering knowledge-based questions about videos}. In \bibinfo{booktitle}{\emph{Proceedings of the AAAI conference on artificial intelligence}}, Vol.~\bibinfo{volume}{34}. \bibinfo{pages}{10826--10834}.
\newblock


\bibitem[Gemini(2023)]%
        {team2023gemini}
\bibfield{author}{\bibinfo{person}{Team Gemini}.} \bibinfo{year}{2023}\natexlab{}.
\newblock \showarticletitle{Gemini: a family of highly capable multimodal models}.
\newblock \bibinfo{journal}{\emph{arXiv preprint arXiv:2312.11805}} (\bibinfo{year}{2023}).
\newblock


\bibitem[Goyal et~al\mbox{.}(2017)]%
        {goyal2017making}
\bibfield{author}{\bibinfo{person}{Yash Goyal}, \bibinfo{person}{Tejas Khot}, \bibinfo{person}{Douglas Summers-Stay}, \bibinfo{person}{Dhruv Batra}, {and} \bibinfo{person}{Devi Parikh}.} \bibinfo{year}{2017}\natexlab{}.
\newblock \showarticletitle{Making the v in vqa matter: Elevating the role of image understanding in visual question answering}. In \bibinfo{booktitle}{\emph{Proceedings of the IEEE conference on computer vision and pattern recognition}}. \bibinfo{pages}{6904--6913}.
\newblock


\bibitem[Gui et~al\mbox{.}(2022)]%
        {kat}
\bibfield{author}{\bibinfo{person}{Liangke Gui}, \bibinfo{person}{Borui Wang}, \bibinfo{person}{Qiuyuan Huang}, \bibinfo{person}{Alexander Hauptmann}, \bibinfo{person}{Yonatan Bisk}, {and} \bibinfo{person}{Jianfeng Gao}.} \bibinfo{year}{2022}\natexlab{}.
\newblock \showarticletitle{{KAT}: A Knowledge Augmented Transformer for Vision-and-Language}. In \bibinfo{booktitle}{\emph{Proceedings of the 2022 Conference of the North American Chapter of the Association for Computational Linguistics: Human Language Technologies}}. \bibinfo{publisher}{Association for Computational Linguistics}, \bibinfo{address}{Seattle, United States}, \bibinfo{pages}{956--968}.
\newblock
\urldef\tempurl%
\url{https://doi.org/10.18653/v1/2022.naacl-main.70}
\showDOI{\tempurl}


\bibitem[Guo et~al\mbox{.}(2022)]%
        {unifer}
\bibfield{author}{\bibinfo{person}{Yangyang Guo}, \bibinfo{person}{Liqiang Nie}, \bibinfo{person}{Yongkang Wong}, \bibinfo{person}{Yibing Liu}, \bibinfo{person}{Zhiyong Cheng}, {and} \bibinfo{person}{Mohan Kankanhalli}.} \bibinfo{year}{2022}\natexlab{}.
\newblock \showarticletitle{A Unified End-to-End Retriever-Reader Framework for Knowledge-Based VQA}. In \bibinfo{booktitle}{\emph{Proceedings of the 30th ACM International Conference on Multimedia}} (Lisboa, Portugal) \emph{(\bibinfo{series}{MM '22})}. \bibinfo{publisher}{Association for Computing Machinery}, \bibinfo{address}{New York, NY, USA}, \bibinfo{pages}{2061–2069}.
\newblock
\showISBNx{9781450392037}
\urldef\tempurl%
\url{https://doi.org/10.1145/3503161.3547870}
\showDOI{\tempurl}


\bibitem[Jang et~al\mbox{.}(2019)]%
        {jang2019video}
\bibfield{author}{\bibinfo{person}{Yunseok Jang}, \bibinfo{person}{Yale Song}, \bibinfo{person}{Chris~Dongjoo Kim}, \bibinfo{person}{Youngjae Yu}, \bibinfo{person}{Youngjin Kim}, {and} \bibinfo{person}{Gunhee Kim}.} \bibinfo{year}{2019}\natexlab{}.
\newblock \showarticletitle{Video question answering with spatio-temporal reasoning}.
\newblock \bibinfo{journal}{\emph{International Journal of Computer Vision}}  \bibinfo{volume}{127} (\bibinfo{year}{2019}), \bibinfo{pages}{1385--1412}.
\newblock


\bibitem[Jang et~al\mbox{.}(2017)]%
        {jang2017tgif}
\bibfield{author}{\bibinfo{person}{Yunseok Jang}, \bibinfo{person}{Yale Song}, \bibinfo{person}{Youngjae Yu}, \bibinfo{person}{Youngjin Kim}, {and} \bibinfo{person}{Gunhee Kim}.} \bibinfo{year}{2017}\natexlab{}.
\newblock \showarticletitle{Tgif-qa: Toward spatio-temporal reasoning in visual question answering}. In \bibinfo{booktitle}{\emph{Proceedings of the IEEE conference on computer vision and pattern recognition}}. \bibinfo{pages}{2758--2766}.
\newblock


\bibitem[Kong and Allan(2013)]%
        {Kong2013}
\bibfield{author}{\bibinfo{person}{Weize Kong} {and} \bibinfo{person}{James Allan}.} \bibinfo{year}{2013}\natexlab{}.
\newblock \showarticletitle{Extracting query facets from search results}. In \bibinfo{booktitle}{\emph{Proceedings of the 36th International ACM SIGIR Conference on Research and Development in Information Retrieval}} (Dublin, Ireland) \emph{(\bibinfo{series}{SIGIR '13})}. \bibinfo{publisher}{Association for Computing Machinery}, \bibinfo{address}{New York, NY, USA}, \bibinfo{pages}{93–102}.
\newblock
\showISBNx{9781450320344}
\urldef\tempurl%
\url{https://doi.org/10.1145/2484028.2484097}
\showDOI{\tempurl}


\bibitem[Lee et~al\mbox{.}(2024)]%
        {lee2024nv}
\bibfield{author}{\bibinfo{person}{Chankyu Lee}, \bibinfo{person}{Rajarshi Roy}, \bibinfo{person}{Mengyao Xu}, \bibinfo{person}{Jonathan Raiman}, \bibinfo{person}{Mohammad Shoeybi}, \bibinfo{person}{Bryan Catanzaro}, {and} \bibinfo{person}{Wei Ping}.} \bibinfo{year}{2024}\natexlab{}.
\newblock \showarticletitle{NV-Embed: Improved Techniques for Training LLMs as Generalist Embedding Models}.
\newblock \bibinfo{journal}{\emph{arXiv preprint arXiv:2405.17428}} (\bibinfo{year}{2024}).
\newblock


\bibitem[Lei et~al\mbox{.}(2018)]%
        {lei2018tvqa}
\bibfield{author}{\bibinfo{person}{Jie Lei}, \bibinfo{person}{Licheng Yu}, \bibinfo{person}{Mohit Bansal}, {and} \bibinfo{person}{Tamara~L Berg}.} \bibinfo{year}{2018}\natexlab{}.
\newblock \showarticletitle{Tvqa: Localized, compositional video question answering}.
\newblock \bibinfo{journal}{\emph{arXiv preprint arXiv:1809.01696}} (\bibinfo{year}{2018}).
\newblock


\bibitem[Lewis et~al\mbox{.}(2020)]%
        {rag}
\bibfield{author}{\bibinfo{person}{Patrick Lewis}, \bibinfo{person}{Ethan Perez}, \bibinfo{person}{Aleksandra Piktus}, \bibinfo{person}{Fabio Petroni}, \bibinfo{person}{Vladimir Karpukhin}, \bibinfo{person}{Naman Goyal}, \bibinfo{person}{Heinrich K\"{u}ttler}, \bibinfo{person}{Mike Lewis}, \bibinfo{person}{Wen-tau Yih}, \bibinfo{person}{Tim Rockt\"{a}schel}, \bibinfo{person}{Sebastian Riedel}, {and} \bibinfo{person}{Douwe Kiela}.} \bibinfo{year}{2020}\natexlab{}.
\newblock \showarticletitle{Retrieval-augmented generation for knowledge-intensive NLP tasks}. In \bibinfo{booktitle}{\emph{Proceedings of the 34th International Conference on Neural Information Processing Systems}} (Vancouver, BC, Canada) \emph{(\bibinfo{series}{NIPS '20})}. \bibinfo{publisher}{Curran Associates Inc.}, \bibinfo{address}{Red Hook, NY, USA}, Article \bibinfo{articleno}{793}, \bibinfo{numpages}{16}~pages.
\newblock
\showISBNx{9781713829546}


\bibitem[Li et~al\mbox{.}(2024)]%
        {li2024making}
\bibfield{author}{\bibinfo{person}{Chaofan Li}, \bibinfo{person}{MingHao Qin}, \bibinfo{person}{Shitao Xiao}, \bibinfo{person}{Jianlyu Chen}, \bibinfo{person}{Kun Luo}, \bibinfo{person}{Yingxia Shao}, \bibinfo{person}{Defu Lian}, {and} \bibinfo{person}{Zheng Liu}.} \bibinfo{year}{2024}\natexlab{}.
\newblock \showarticletitle{Making text embedders few-shot learners}.
\newblock \bibinfo{journal}{\emph{arXiv preprint arXiv:2409.15700}} (\bibinfo{year}{2024}).
\newblock


\bibitem[Li et~al\mbox{.}(2023)]%
        {li2023videochat}
\bibfield{author}{\bibinfo{person}{KunChang Li}, \bibinfo{person}{Yinan He}, \bibinfo{person}{Yi Wang}, \bibinfo{person}{Yizhuo Li}, \bibinfo{person}{Wenhai Wang}, \bibinfo{person}{Ping Luo}, \bibinfo{person}{Yali Wang}, \bibinfo{person}{Limin Wang}, {and} \bibinfo{person}{Yu Qiao}.} \bibinfo{year}{2023}\natexlab{}.
\newblock \showarticletitle{Videochat: Chat-centric video understanding}.
\newblock \bibinfo{journal}{\emph{arXiv preprint arXiv:2305.06355}} (\bibinfo{year}{2023}).
\newblock


\bibitem[Li et~al\mbox{.}(2025)]%
        {li2025llama}
\bibfield{author}{\bibinfo{person}{Yanwei Li}, \bibinfo{person}{Chengyao Wang}, {and} \bibinfo{person}{Jiaya Jia}.} \bibinfo{year}{2025}\natexlab{}.
\newblock \showarticletitle{Llama-vid: An image is worth 2 tokens in large language models}. In \bibinfo{booktitle}{\emph{European Conference on Computer Vision}}. Springer, \bibinfo{pages}{323--340}.
\newblock


\bibitem[Lin et~al\mbox{.}(2023)]%
        {lin2023video}
\bibfield{author}{\bibinfo{person}{Bin Lin}, \bibinfo{person}{Yang Ye}, \bibinfo{person}{Bin Zhu}, \bibinfo{person}{Jiaxi Cui}, \bibinfo{person}{Munan Ning}, \bibinfo{person}{Peng Jin}, {and} \bibinfo{person}{Li Yuan}.} \bibinfo{year}{2023}\natexlab{}.
\newblock \showarticletitle{Video-llava: Learning united visual representation by alignment before projection}.
\newblock \bibinfo{journal}{\emph{arXiv preprint arXiv:2311.10122}} (\bibinfo{year}{2023}).
\newblock


\bibitem[Lin(2004)]%
        {lin-2004-rouge}
\bibfield{author}{\bibinfo{person}{Chin-Yew Lin}.} \bibinfo{year}{2004}\natexlab{}.
\newblock \showarticletitle{{ROUGE}: A Package for Automatic Evaluation of Summaries}. In \bibinfo{booktitle}{\emph{Text Summarization Branches Out}}. \bibinfo{publisher}{Association for Computational Linguistics}, \bibinfo{address}{Barcelona, Spain}, \bibinfo{pages}{74--81}.
\newblock
\urldef\tempurl%
\url{https://aclanthology.org/W04-1013/}
\showURL{%
\tempurl}


\bibitem[Loshchilov(2017)]%
        {loshchilov2017decoupled}
\bibfield{author}{\bibinfo{person}{I Loshchilov}.} \bibinfo{year}{2017}\natexlab{}.
\newblock \showarticletitle{Decoupled weight decay regularization}.
\newblock \bibinfo{journal}{\emph{arXiv preprint arXiv:1711.05101}} (\bibinfo{year}{2017}).
\newblock


\bibitem[Lu et~al\mbox{.}(2018)]%
        {lu2018r}
\bibfield{author}{\bibinfo{person}{Pan Lu}, \bibinfo{person}{Lei Ji}, \bibinfo{person}{Wei Zhang}, \bibinfo{person}{Nan Duan}, \bibinfo{person}{Ming Zhou}, {and} \bibinfo{person}{Jianyong Wang}.} \bibinfo{year}{2018}\natexlab{}.
\newblock \showarticletitle{R-VQA: learning visual relation facts with semantic attention for visual question answering}. In \bibinfo{booktitle}{\emph{Proceedings of the 24th ACM SIGKDD International Conference on Knowledge Discovery \& Data Mining}}. \bibinfo{pages}{1880--1889}.
\newblock


\bibitem[Marino et~al\mbox{.}(2020)]%
        {krisp}
\bibfield{author}{\bibinfo{person}{Kenneth Marino}, \bibinfo{person}{Xinlei Chen}, \bibinfo{person}{Devi Parikh}, \bibinfo{person}{Abhinav~Kumar Gupta}, {and} \bibinfo{person}{Marcus Rohrbach}.} \bibinfo{year}{2020}\natexlab{}.
\newblock \showarticletitle{KRISP: Integrating Implicit and Symbolic Knowledge for Open-Domain Knowledge-Based VQA}. In \bibinfo{booktitle}{\emph{2021 IEEE/CVF Conference on Computer Vision and Pattern Recognition (CVPR)}}. \bibinfo{pages}{14106--14116}.
\newblock


\bibitem[Marino et~al\mbox{.}(2019)]%
        {marino2019ok}
\bibfield{author}{\bibinfo{person}{Kenneth Marino}, \bibinfo{person}{Mohammad Rastegari}, \bibinfo{person}{Ali Farhadi}, {and} \bibinfo{person}{Roozbeh Mottaghi}.} \bibinfo{year}{2019}\natexlab{}.
\newblock \showarticletitle{Ok-vqa: A visual question answering benchmark requiring external knowledge}. In \bibinfo{booktitle}{\emph{Proceedings of the IEEE/cvf conference on computer vision and pattern recognition}}. \bibinfo{pages}{3195--3204}.
\newblock


\bibitem[Muennighoff et~al\mbox{.}(2023)]%
        {muennighoff-etal-2023-mteb}
\bibfield{author}{\bibinfo{person}{Niklas Muennighoff}, \bibinfo{person}{Nouamane Tazi}, \bibinfo{person}{Loic Magne}, {and} \bibinfo{person}{Nils Reimers}.} \bibinfo{year}{2023}\natexlab{}.
\newblock \showarticletitle{{MTEB}: Massive Text Embedding Benchmark}. In \bibinfo{booktitle}{\emph{Proceedings of the 17th Conference of the European Chapter of the Association for Computational Linguistics}}, \bibfield{editor}{\bibinfo{person}{Andreas Vlachos} {and} \bibinfo{person}{Isabelle Augenstein}} (Eds.). \bibinfo{publisher}{Association for Computational Linguistics}, \bibinfo{address}{Dubrovnik, Croatia}, \bibinfo{pages}{2014--2037}.
\newblock
\urldef\tempurl%
\url{https://doi.org/10.18653/v1/2023.eacl-main.148}
\showDOI{\tempurl}


\bibitem[OpenAI(2023)]%
        {achiam2023gpt}
\bibfield{author}{\bibinfo{person}{OpenAI}.} \bibinfo{year}{2023}\natexlab{}.
\newblock \showarticletitle{Gpt-4 technical report}.
\newblock \bibinfo{journal}{\emph{arXiv preprint arXiv:2303.08774}} (\bibinfo{year}{2023}).
\newblock


\bibitem[Papineni et~al\mbox{.}(2002)]%
        {papineni2002bleu}
\bibfield{author}{\bibinfo{person}{Kishore Papineni}, \bibinfo{person}{Salim Roukos}, \bibinfo{person}{Todd Ward}, {and} \bibinfo{person}{Wei-Jing Zhu}.} \bibinfo{year}{2002}\natexlab{}.
\newblock \showarticletitle{Bleu: a method for automatic evaluation of machine translation}. In \bibinfo{booktitle}{\emph{Proceedings of the 40th annual meeting of the Association for Computational Linguistics}}. \bibinfo{pages}{311--318}.
\newblock


\bibitem[Qu et~al\mbox{.}(2021)]%
        {Qu2021KIVQA}
\bibfield{author}{\bibinfo{person}{Chen Qu}, \bibinfo{person}{Hamed Zamani}, \bibinfo{person}{Liu Yang}, \bibinfo{person}{W.~Bruce Croft}, {and} \bibinfo{person}{Erik Learned-Miller}.} \bibinfo{year}{2021}\natexlab{}.
\newblock \showarticletitle{Passage Retrieval for Outside-Knowledge Visual Question Answering}. In \bibinfo{booktitle}{\emph{Proceedings of the 44th International ACM SIGIR Conference on Research and Development in Information Retrieval}} (Virtual Event, Canada) \emph{(\bibinfo{series}{SIGIR '21})}. \bibinfo{publisher}{Association for Computing Machinery}, \bibinfo{address}{New York, NY, USA}, \bibinfo{pages}{1753–1757}.
\newblock
\showISBNx{9781450380379}
\urldef\tempurl%
\url{https://doi.org/10.1145/3404835.3462987}
\showDOI{\tempurl}


\bibitem[Radford et~al\mbox{.}(2021)]%
        {radford2021learning}
\bibfield{author}{\bibinfo{person}{Alec Radford} {et~al\mbox{.}}} \bibinfo{year}{2021}\natexlab{}.
\newblock \showarticletitle{Learning transferable visual models from natural language supervision}. In \bibinfo{booktitle}{\emph{International conference on machine learning}}. PMLR, \bibinfo{pages}{8748--8763}.
\newblock


\bibitem[Robertson and Walker(1994)]%
        {robertson1994some}
\bibfield{author}{\bibinfo{person}{Stephen~E Robertson} {and} \bibinfo{person}{Steve Walker}.} \bibinfo{year}{1994}\natexlab{}.
\newblock \showarticletitle{Some simple effective approximations to the 2-poisson model for probabilistic weighted retrieval}. In \bibinfo{booktitle}{\emph{SIGIR’94: Proceedings of the Seventeenth Annual International ACM-SIGIR Conference on Research and Development in Information Retrieval, organised by Dublin City University}}. Springer, \bibinfo{pages}{232--241}.
\newblock


\bibitem[Salemi et~al\mbox{.}(2023a)]%
        {Salemi2023MMFID}
\bibfield{author}{\bibinfo{person}{Alireza Salemi}, \bibinfo{person}{Juan Altmayer~Pizzorno}, {and} \bibinfo{person}{Hamed Zamani}.} \bibinfo{year}{2023}\natexlab{a}.
\newblock \showarticletitle{A Symmetric Dual Encoding Dense Retrieval Framework for Knowledge-Intensive Visual Question Answering}. In \bibinfo{booktitle}{\emph{Proceedings of the 46th International ACM SIGIR Conference on Research and Development in Information Retrieval}} (Taipei, Taiwan) \emph{(\bibinfo{series}{SIGIR '23})}. \bibinfo{publisher}{Association for Computing Machinery}, \bibinfo{address}{New York, NY, USA}, \bibinfo{pages}{110–120}.
\newblock
\showISBNx{9781450394086}
\urldef\tempurl%
\url{https://doi.org/10.1145/3539618.3591629}
\showDOI{\tempurl}


\bibitem[Salemi et~al\mbox{.}(2023b)]%
        {Salemi2023ICTIR}
\bibfield{author}{\bibinfo{person}{Alireza Salemi}, \bibinfo{person}{Mahta Rafiee}, {and} \bibinfo{person}{Hamed Zamani}.} \bibinfo{year}{2023}\natexlab{b}.
\newblock \showarticletitle{Pre-Training Multi-Modal Dense Retrievers for Outside-Knowledge Visual Question Answering}. In \bibinfo{booktitle}{\emph{Proceedings of the 2023 ACM SIGIR International Conference on Theory of Information Retrieval}} (Taipei, Taiwan) \emph{(\bibinfo{series}{ICTIR '23})}. \bibinfo{publisher}{Association for Computing Machinery}, \bibinfo{address}{New York, NY, USA}, \bibinfo{pages}{169–176}.
\newblock
\showISBNx{9798400700736}
\urldef\tempurl%
\url{https://doi.org/10.1145/3578337.3605137}
\showDOI{\tempurl}


\bibitem[Sun et~al\mbox{.}(2023)]%
        {sun2023eva}
\bibfield{author}{\bibinfo{person}{Quan Sun}, \bibinfo{person}{Yuxin Fang}, \bibinfo{person}{Ledell Wu}, \bibinfo{person}{Xinlong Wang}, {and} \bibinfo{person}{Yue Cao}.} \bibinfo{year}{2023}\natexlab{}.
\newblock \showarticletitle{Eva-clip: Improved training techniques for clip at scale}.
\newblock \bibinfo{journal}{\emph{arXiv preprint arXiv:2303.15389}} (\bibinfo{year}{2023}).
\newblock


\bibitem[Touvron et~al\mbox{.}(2023)]%
        {touvron2023llama}
\bibfield{author}{\bibinfo{person}{Hugo Touvron}, \bibinfo{person}{Thibaut Lavril}, \bibinfo{person}{Gautier Izacard}, \bibinfo{person}{Xavier Martinet}, \bibinfo{person}{Marie-Anne Lachaux}, \bibinfo{person}{Timoth{\'e}e Lacroix}, \bibinfo{person}{Baptiste Rozi{\`e}re}, \bibinfo{person}{Naman Goyal}, \bibinfo{person}{Eric Hambro}, \bibinfo{person}{Faisal Azhar}, {et~al\mbox{.}}} \bibinfo{year}{2023}\natexlab{}.
\newblock \showarticletitle{Llama: Open and efficient foundation language models}.
\newblock \bibinfo{journal}{\emph{arXiv preprint arXiv:2302.13971}} (\bibinfo{year}{2023}).
\newblock


\bibitem[Wang et~al\mbox{.}(2022)]%
        {wang2022text}
\bibfield{author}{\bibinfo{person}{Liang Wang}, \bibinfo{person}{Nan Yang}, \bibinfo{person}{Xiaolong Huang}, \bibinfo{person}{Binxing Jiao}, \bibinfo{person}{Linjun Yang}, \bibinfo{person}{Daxin Jiang}, \bibinfo{person}{Rangan Majumder}, {and} \bibinfo{person}{Furu Wei}.} \bibinfo{year}{2022}\natexlab{}.
\newblock \showarticletitle{Text embeddings by weakly-supervised contrastive pre-training}.
\newblock \bibinfo{journal}{\emph{arXiv preprint arXiv:2212.03533}} (\bibinfo{year}{2022}).
\newblock


\bibitem[Wang et~al\mbox{.}(2023)]%
        {wang2023improving}
\bibfield{author}{\bibinfo{person}{Liang Wang}, \bibinfo{person}{Nan Yang}, \bibinfo{person}{Xiaolong Huang}, \bibinfo{person}{Linjun Yang}, \bibinfo{person}{Rangan Majumder}, {and} \bibinfo{person}{Furu Wei}.} \bibinfo{year}{2023}\natexlab{}.
\newblock \showarticletitle{Improving text embeddings with large language models}.
\newblock \bibinfo{journal}{\emph{arXiv preprint arXiv:2401.00368}} (\bibinfo{year}{2023}).
\newblock


\bibitem[Wang et~al\mbox{.}(2024c)]%
        {wang2024multilingual}
\bibfield{author}{\bibinfo{person}{Liang Wang}, \bibinfo{person}{Nan Yang}, \bibinfo{person}{Xiaolong Huang}, \bibinfo{person}{Linjun Yang}, \bibinfo{person}{Rangan Majumder}, {and} \bibinfo{person}{Furu Wei}.} \bibinfo{year}{2024}\natexlab{c}.
\newblock \showarticletitle{Multilingual e5 text embeddings: A technical report}.
\newblock \bibinfo{journal}{\emph{arXiv preprint arXiv:2402.05672}} (\bibinfo{year}{2024}).
\newblock


\bibitem[Wang et~al\mbox{.}(2024a)]%
        {wang2024qwen2vl}
\bibfield{author}{\bibinfo{person}{Peng Wang}, \bibinfo{person}{Shuai Bai}, \bibinfo{person}{Sinan Tan}, \bibinfo{person}{Shijie Wang}, \bibinfo{person}{Zhihao Fan}, \bibinfo{person}{Jinze Bai}, \bibinfo{person}{Keqin Chen}, \bibinfo{person}{Xuejing Liu}, \bibinfo{person}{Jialin Wang}, \bibinfo{person}{Wenbin Ge}, \bibinfo{person}{Yang Fan}, \bibinfo{person}{Kai Dang}, \bibinfo{person}{Mengfei Du}, \bibinfo{person}{Xuancheng Ren}, \bibinfo{person}{Rui Men}, \bibinfo{person}{Dayiheng Liu}, \bibinfo{person}{Chang Zhou}, \bibinfo{person}{Jingren Zhou}, {and} \bibinfo{person}{Junyang Lin}.} \bibinfo{year}{2024}\natexlab{a}.
\newblock \bibinfo{title}{Qwen2-VL: Enhancing Vision-Language Model's Perception of the World at Any Resolution}.
\newblock
\newblock
\showeprint[arxiv]{2409.12191}~[cs.CV]
\urldef\tempurl%
\url{https://arxiv.org/abs/2409.12191}
\showURL{%
\tempurl}


\bibitem[Wang et~al\mbox{.}(2017)]%
        {wang2017fvqa}
\bibfield{author}{\bibinfo{person}{Peng Wang}, \bibinfo{person}{Qi Wu}, \bibinfo{person}{Chunhua Shen}, \bibinfo{person}{Anthony Dick}, {and} \bibinfo{person}{Anton Van Den~Hengel}.} \bibinfo{year}{2017}\natexlab{}.
\newblock \showarticletitle{Fvqa: Fact-based visual question answering}.
\newblock \bibinfo{journal}{\emph{IEEE transactions on pattern analysis and machine intelligence}} \bibinfo{volume}{40}, \bibinfo{number}{10} (\bibinfo{year}{2017}), \bibinfo{pages}{2413--2427}.
\newblock


\bibitem[Wang et~al\mbox{.}(2015)]%
        {wang2015explicit}
\bibfield{author}{\bibinfo{person}{Peng Wang}, \bibinfo{person}{Qi Wu}, \bibinfo{person}{Chunhua Shen}, \bibinfo{person}{Anton van~den Hengel}, {and} \bibinfo{person}{Anthony Dick}.} \bibinfo{year}{2015}\natexlab{}.
\newblock \showarticletitle{Explicit knowledge-based reasoning for visual question answering}.
\newblock \bibinfo{journal}{\emph{arXiv preprint arXiv:1511.02570}} (\bibinfo{year}{2015}).
\newblock


\bibitem[Wang et~al\mbox{.}(2024b)]%
        {wang2024intern2vl}
\bibfield{author}{\bibinfo{person}{Weiyun Wang}, \bibinfo{person}{Zhe Chen}, \bibinfo{person}{Wenhai Wang}, \bibinfo{person}{Yue Cao}, \bibinfo{person}{Yangzhou Liu}, \bibinfo{person}{Zhangwei Gao}, \bibinfo{person}{Jinguo Zhu}, \bibinfo{person}{Xizhou Zhu}, \bibinfo{person}{Lewei Lu}, \bibinfo{person}{Yu Qiao}, {and} \bibinfo{person}{Jifeng Dai}.} \bibinfo{year}{2024}\natexlab{b}.
\newblock \bibinfo{title}{Enhancing the Reasoning Ability of Multimodal Large Language Models via Mixed Preference Optimization}.
\newblock
\newblock
\showeprint[arxiv]{2411.10442}~[cs.CL]
\urldef\tempurl%
\url{https://arxiv.org/abs/2411.10442}
\showURL{%
\tempurl}


\bibitem[Wu et~al\mbox{.}(2022)]%
        {mavex}
\bibfield{author}{\bibinfo{person}{Jialin Wu}, \bibinfo{person}{Jiasen Lu}, \bibinfo{person}{Ashish Sabharwal}, {and} \bibinfo{person}{Roozbeh Mottaghi}.} \bibinfo{year}{2022}\natexlab{}.
\newblock \showarticletitle{Multi-Modal Answer Validation for Knowledge-Based VQA}. In \bibinfo{booktitle}{\emph{Proceedings of the AAAI Conference on Artificial Intelligence}}. \bibinfo{pages}{2712--2721}.
\newblock
\urldef\tempurl%
\url{https://doi.org/10.1609/aaai.v36i3.20174}
\showDOI{\tempurl}


\bibitem[Wu et~al\mbox{.}(2016)]%
        {wu2016ask}
\bibfield{author}{\bibinfo{person}{Qi Wu}, \bibinfo{person}{Peng Wang}, \bibinfo{person}{Chunhua Shen}, \bibinfo{person}{Anthony Dick}, {and} \bibinfo{person}{Anton Van Den~Hengel}.} \bibinfo{year}{2016}\natexlab{}.
\newblock \showarticletitle{Ask me anything: Free-form visual question answering based on knowledge from external sources}. In \bibinfo{booktitle}{\emph{Proceedings of the IEEE conference on computer vision and pattern recognition}}. \bibinfo{pages}{4622--4630}.
\newblock


\bibitem[Wu et~al\mbox{.}(2021)]%
        {wu2021transferring}
\bibfield{author}{\bibinfo{person}{Tianran Wu}, \bibinfo{person}{Noa Garcia}, \bibinfo{person}{Mayu Otani}, \bibinfo{person}{Chenhui Chu}, \bibinfo{person}{Yuta Nakashima}, {and} \bibinfo{person}{Haruo Takemura}.} \bibinfo{year}{2021}\natexlab{}.
\newblock \showarticletitle{Transferring domain-agnostic knowledge in video question answering}.
\newblock \bibinfo{journal}{\emph{arXiv preprint arXiv:2110.13395}} (\bibinfo{year}{2021}).
\newblock


\bibitem[Xiao et~al\mbox{.}(2024a)]%
        {xiao2024videoqa}
\bibfield{author}{\bibinfo{person}{Junbin Xiao}, \bibinfo{person}{Nanxin Huang}, \bibinfo{person}{Hangyu Qin}, \bibinfo{person}{Dongyang Li}, \bibinfo{person}{Yicong Li}, \bibinfo{person}{Fengbin Zhu}, \bibinfo{person}{Zhulin Tao}, \bibinfo{person}{Jianxing Yu}, \bibinfo{person}{Liang Lin}, \bibinfo{person}{Tat-Seng Chua}, {et~al\mbox{.}}} \bibinfo{year}{2024}\natexlab{a}.
\newblock \showarticletitle{VideoQA in the Era of LLMs: An Empirical Study}.
\newblock \bibinfo{journal}{\emph{arXiv preprint arXiv:2408.04223}} (\bibinfo{year}{2024}).
\newblock


\bibitem[Xiao et~al\mbox{.}(2024b)]%
        {xiao2024can}
\bibfield{author}{\bibinfo{person}{Junbin Xiao}, \bibinfo{person}{Angela Yao}, \bibinfo{person}{Yicong Li}, {and} \bibinfo{person}{Tat-Seng Chua}.} \bibinfo{year}{2024}\natexlab{b}.
\newblock \showarticletitle{Can i trust your answer? visually grounded video question answering}. In \bibinfo{booktitle}{\emph{Proceedings of the IEEE/CVF Conference on Computer Vision and Pattern Recognition}}. \bibinfo{pages}{13204--13214}.
\newblock


\bibitem[Xu et~al\mbox{.}(2017)]%
        {xu2017video}
\bibfield{author}{\bibinfo{person}{Dejing Xu}, \bibinfo{person}{Zhou Zhao}, \bibinfo{person}{Jun Xiao}, \bibinfo{person}{Fei Wu}, \bibinfo{person}{Hanwang Zhang}, \bibinfo{person}{Xiangnan He}, {and} \bibinfo{person}{Yueting Zhuang}.} \bibinfo{year}{2017}\natexlab{}.
\newblock \showarticletitle{Video question answering via gradually refined attention over appearance and motion}. In \bibinfo{booktitle}{\emph{Proceedings of the 25th ACM international conference on Multimedia}}. \bibinfo{pages}{1645--1653}.
\newblock


\bibitem[Zamani et~al\mbox{.}(2022)]%
        {reml}
\bibfield{author}{\bibinfo{person}{Hamed Zamani}, \bibinfo{person}{Fernando Diaz}, \bibinfo{person}{Mostafa Dehghani}, \bibinfo{person}{Donald Metzler}, {and} \bibinfo{person}{Michael Bendersky}.} \bibinfo{year}{2022}\natexlab{}.
\newblock \showarticletitle{Retrieval-Enhanced Machine Learning}. In \bibinfo{booktitle}{\emph{Proceedings of the 45th International ACM SIGIR Conference on Research and Development in Information Retrieval}} (Madrid, Spain) \emph{(\bibinfo{series}{SIGIR '22})}. \bibinfo{publisher}{Association for Computing Machinery}, \bibinfo{address}{New York, NY, USA}, \bibinfo{pages}{2875–2886}.
\newblock
\showISBNx{9781450387323}
\urldef\tempurl%
\url{https://doi.org/10.1145/3477495.3531722}
\showDOI{\tempurl}


\bibitem[Zhang et~al\mbox{.}(2024)]%
        {zhang2024jasper}
\bibfield{author}{\bibinfo{person}{Dun Zhang} {et~al\mbox{.}}} \bibinfo{year}{2024}\natexlab{}.
\newblock \showarticletitle{Jasper and Stella: distillation of SOTA embedding models}.
\newblock \bibinfo{journal}{\emph{arXiv preprint arXiv:2412.19048}} (\bibinfo{year}{2024}).
\newblock


\bibitem[Zhang et~al\mbox{.}(2019)]%
        {zhang2019bertscore}
\bibfield{author}{\bibinfo{person}{Tianyi Zhang}, \bibinfo{person}{Varsha Kishore}, \bibinfo{person}{Felix Wu}, \bibinfo{person}{Kilian~Q Weinberger}, {and} \bibinfo{person}{Yoav Artzi}.} \bibinfo{year}{2019}\natexlab{}.
\newblock \showarticletitle{Bertscore: Evaluating text generation with bert}.
\newblock \bibinfo{journal}{\emph{arXiv preprint arXiv:1904.09675}} (\bibinfo{year}{2019}).
\newblock


\bibitem[Zhu et~al\mbox{.}(2015)]%
        {zhu2015building}
\bibfield{author}{\bibinfo{person}{Yuke Zhu}, \bibinfo{person}{Ce Zhang}, \bibinfo{person}{Christopher R{\'e}}, {and} \bibinfo{person}{Li Fei-Fei}.} \bibinfo{year}{2015}\natexlab{}.
\newblock \showarticletitle{Building a large-scale multimodal knowledge base system for answering visual queries}.
\newblock \bibinfo{journal}{\emph{arXiv preprint arXiv:1507.05670}} (\bibinfo{year}{2015}).
\newblock


\end{thebibliography}
